\documentclass[reprint,amsfonts,amsmath,amssymb,jcp,aip]{revtex4-2}

\usepackage{graphicx}
\usepackage{dcolumn}
\usepackage{bm}
\usepackage{hyperref}
\usepackage[final]{changes}
\usepackage{dsfont}
\usepackage{nccmath,xfrac}       
\usepackage{mathtools}
\usepackage{lineno}
\usepackage{float}
\usepackage{placeins}
\usepackage[T1]{fontenc}  
\usepackage[utf8]{inputenc} 
\def\sep{\,\vert\,}
\graphicspath{{../figs}}

\begin{document}

\title{Exclusion and internal structure in the association/dissociation
  kinetics of DNA-binding proteins in the fast diffusion limit}


\author{Xiangting Li}
\affiliation{%
  School of Mathematics and Statistics, Southwest University,
  Chongqing, PRC
}%

\author{Tom Chou}%
 \email{tomchou@ucla.edu}
\affiliation{%
Department of Biomathematics, University of California, Los Angeles, CA 90095, USA.
}%
\date{\today}

\begin{abstract}
In many canonical biophysical and biochemical contexts, complex
diffusing DNA-binding proteins undergoing nonspecific adsorption and
desorption can exist in different conformational states.  Equilibrium
binding theories and kinetic models of competing ligands provide a
basis for describing these systems. Coupling the internal
conformational kinetics of bound proteins to exclusion and diffusion
requires tracking both protein positions and binding
states. Closed-form, tractable mathematical models that accurately
describe the stochastic dynamics of structured proteins that bind to
and diffuse along DNA are lacking. Here, we develop a new
coarse-grained stochastic model of nonspecific DNA-protein binding in
the fast diffusion regime, from which we derive an effective Markov
model and its corresponding mean-field approximation that incorporates
protein footprint size and conformational switching.  We fit our new
model quantitatively to previously measured dynamics of Replication
Protein A (RPA) binding to ssDNA. Finally, by including multisegment
binding states, our analysis captures facilitated exchange, revealing
how multisegment architecture balances high affinity with rapid
remodeling. Our theoretical model applies broadly to many other
DNA-binding protein systems, including nucleosomes, ssDNA-binding
proteins, and other multisegment complexes, providing accurate
predictions and new strategies for evaluating experimental data.
\end{abstract}

\maketitle
\section{Introduction}
\label{sec:introduction}

DNA-binding proteins in crowded genomes are crucial players in
regulation, differentiation, and replication. The kinetics of
DNA-binding proteins fall into two broad classes: fast-diffusing
``searchers,'' which slide along the helix to locate specific target
sites \cite{Halford2004June,Belotserkovskii2021February}, and
fast-diffusing ``occluders,'' such as histones, replication protein A
(RPA), and Single-Strand Binding protein (SSB), which rapidly
redistribute but assemble into high-affinity filaments that shield DNA
from other factors \cite{Lohman1985March,Fanning2006August}. Occluders
need to slide \cite{histones_prl,Bowman2010,Armeev2021}, redistribute,
and be displaced to ensure DNA accessibility for subsequent processes
\cite{Nguyen2014September,Chen2016April}. Further complexity arises as
many such proteins, including histones, RPA, and SSB, bind DNA in
\textit{multi-conformational} modes with varying footprints
\cite{Lohman1985March,histones_2003,Fanning2006August,Pokhrel2019February,
  Li2023November}. The combination of rapid diffusion (often operating
in a ``fast diffusion limit'' where positional equilibration is much
faster than that of binding/unbinding \cite{Nguyen2014September}) and
protein conformational \added{changes} creates highly dynamic
protein-DNA assemblies. Accurately modeling the collective kinetics of
these systems---predicting how factors like protein concentration,
binding modes, and diffusion influence phenomena like protein exchange
and DNA site accessibility---remains a major \deleted{modeling}
challenge.

Conventional approaches to modeling DNA-binding protein kinetics often
rely on full lattice models, representing DNA as a grid and tracking
each protein's position and state
\cite{McGhee1974June,Ding2023June}. However, the combinatorial
complexity inherent in such models is immense. Accounting for multiple
proteins, their potential locations, exclusion effects, and multiple
binding conformations leads to an explosion in the number of possible
microscopic states. This renders brute-force stochastic simulations
computationally prohibitive for large DNA lengths, high protein
densities, or extensive parameter exploration, especially when trying
to capture behavior over biologically relevant timescales
\cite{Ding2023June}. Thus, there is a pressing need for reduced
descriptions that capture the essential dynamics of
multi-conformational, fast-diffusing DNA-binding proteins without the
computational burden of tracking every microscopic detail.

McGhee and von Hippel derived the equilibrium statistics of large
ligands binding to a linear polymer \cite{McGhee1974June}. It is known
that the \deleted{distribution of} inter-ligand distances on a linear
polymer follow a geometric distribution. A related model, the Tonks
gas \cite{Tonks1936November}, was developed to quantify the volume
exclusion effect of hard elastic spheres of finite size in a gas-like
system. \added{Hill's equilibrium treatment also includes extended,
directional ligands and ligand-induced conformational changes of the
underlying lattice \cite[Ch.~8]{hill2013}.}
Villaluenga et al. \cite{Villaluenga2020July} showed that the
1D Tonks gas model and the McGhee-von Hippel model are equivalent
\added{at} thermodynamic equilibrium and derived the deterministic mean-field
kinetic equation for a single finite-sized ligand binding to a linear
polymer \replaced{without explicitly modeling diffusion along the
chain}{in the fast diffusion limit}, \replaced{refining the count of
accessible binding positions relative to the earlier coarse-grained
approximation \cite{Jarillo2017April}}{correcting a prior oversimplified
model \cite{Jarillo2017April} where the volume exclusion effect was
not considered}. \added{Villaluenga et al. subsequently
derived kinetic equations for competing ligand types with fixed
footprints, including cooperative binding \cite{villa2023}. Their
equations omit conversion between bound modes, which they identify as
requiring further analysis.}

While the McGhee-von Hippel-Tonks model (\added{hereafter,} the Tonks
gas model\deleted{in the following}) described in
Ref.~[\onlinecite{Villaluenga2020July}] \replaced{and its competitive
  generalization \cite{villa2023} treat attachment and detachment of
  ligands with prescribed footprints, we also track footprint-changing
  transitions of a protein that remains bound to the lattice. Such
  transitions depend on the adjacent gap and couple internal
  conformational kinetics to local exclusion, and arise in processes
  such as nucleosome formation and remodeling
  \cite{histones_2003,histones_prl,Bowman2010,Armeev2021}.}{provides a
  satisfactory framework for the kinetics of a single ligand binding
  to a linear polymer in the fast diffusion limit, it is not directly
  applicable to the kinetics of multiple-ligand binding or internal
  conformations within a single ligand.} \added{For a single bound
  state, with conversion disabled and $\rho_2=0$, writing the coverage
  as $c=a_1\rho_1$ reduces our mean-field equation to Eq.~(7) of
  Ref.~[\onlinecite{Villaluenga2020July}]. Adjustable-length Tonks gas
  models also account for footprint flexibility in
  \textit{equilibrium} adsorption isotherms \cite{histones_2003}.}
Subsequent studies into the conformations of E. coli SSB binding to
DNA \cite{Naufer2021jan} continue to use the simple model in
Refs.~[\onlinecite{Jarillo2017April}] and [\onlinecite{Morin2017May}]
\added{even though its predictions agree} with those of the Tonks gas
model only in the low coverage limit.

Since the diffusion coefficient $D$ has different units than binding
and unbinding rates, it is not clear how fast the diffusion needs to
be in order for the fast diffusion limit in the Tonks gas model to
hold. Lastly, the Tonks gas model assumes that the overall binding
rate scales directly with the number of available binding positions.
As a result, \added{a protein binds to a gap exactly the size of its
  footprint at only half the rate at which it binds to a gap one
  nucleotide larger.} The combinatorial dependence of molecular
binding rates on discrete-length gaps arises in processes such as
random sequential adsorption \cite{Boucher,GAP}.  In reality, proteins
likely sample transient intermediates with smaller footprints when
scanning DNA, so binding rates to gaps of similar size should be
nearly the same.

In this work, we address these challenges by deriving and analyzing a
simplified stochastic model specifically tailored for
multi-conformational DNA-protein binding in the fast diffusion limit.


Following a \added{combinatoricial} approach \added{similar to that}
in Ref.~[\onlinecite{McGhee1974June}], we develop a reduced-state continuous-time
Markov chain using the kinetics of RPA binding to ssDNA as an
example. This model aggregates positionally distinct microstates into
effective states characterized by the number and conformation of bound
proteins, dramatically reducing the state space. We further derive a
corresponding hydrodynamic limit as a system of ordinary differential
equations (ODEs). To validate our approach, we compare the predictions
of the reduced model and its mean-field approximation against detailed
agent-based simulations of the full lattice system and against
experimental data on RPA binding dynamics.

Using this validated framework, we investigate how the interplay between
binding, multi-conformational switching, and fast diffusion governs DNA
accessibility and protein turnover. We focus on RPA as a case study,
quantifying how factors like protein concentration and binding mode
preferences influence the fraction of time DNA sites remain accessible
and the rate of protein exchange. Later on, we generalize the framework
to multi-conformational and multi-step binding of DNA-protein complexes.
Our analysis elucidates how the ability of proteins to adopt different
footprints modulates competitive interactions on the DNA. Ultimately,
this work provides both a powerful modeling methodology and insights
into how dynamic, multi-modal protein-DNA interactions regulate access
to the genome.

\section{Methods}
\begin{figure*}[htbp!]
  \centering
  \includegraphics[width=5.1in]{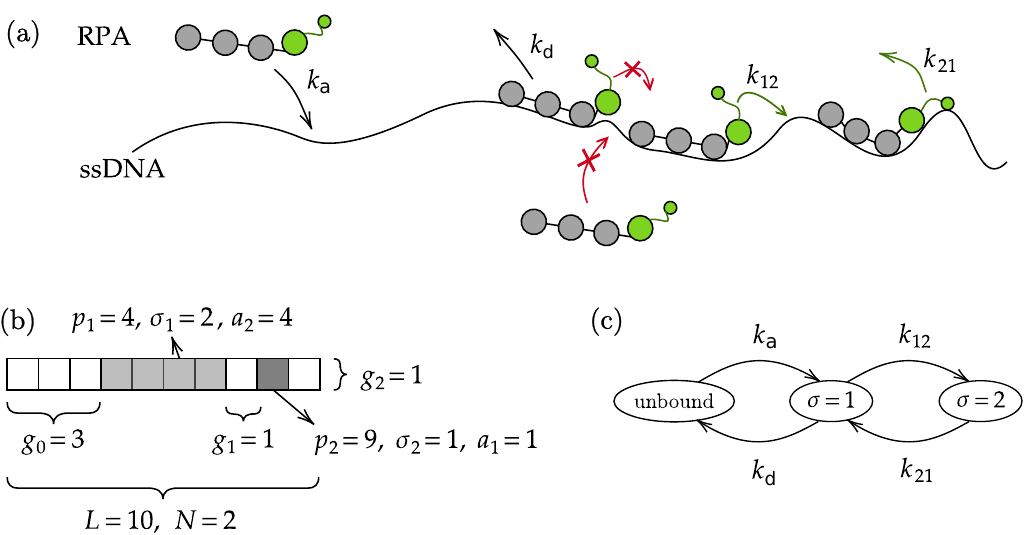}
  \caption{Schematic of the kinetic model of RPA binding to ssDNA
    under volume exclusion. (a) The binding of a protein (RPA) to DNA
    (ssDNA) is subject to volume exclusion constraints. Black arrows
    indicate attachment and detachment of RPA through the binding and
    unbinding of the gray domains (footprint $a_1$) to ssDNA. Red
    arrows indicate prohibited transitions due to volume exclusion
    while green arrows indicate allowed conformation-change
    transitions in RPA.  (b) Schematic representation of the
    microstate of a system with $L=10$, $N=2$, $a_1=1$, $a_2=4$. (c)
    Schematic of the possible transitions within single RPA molecule
    states without volume exclusion constraints.}
  \label{fig:rpa-schematics-1}
\end{figure*}

\subsection{Kinetic model of RPA binding to ssDNA}

%
%
For concreteness, we base our initial derivations primarily on the
RPA-ssDNA binding \added{system} depicted in
Fig.~\ref{fig:rpa-schematics-1}(a) as \added{it} contains all
properties that our model seeks to quantify. The models we develop can
easily be modified and applied to other DNA-protein binding
systems. To formally represent the volume-excluding kinetics of
protein binding, we consider a one-dimensional DNA lattice with $L$
sites (nucleotides), each of which can be either empty or occluded by
one of the $N$ proteins bound to the DNA.

In the context of RPA binding to ssDNA, bound proteins can be in two
different conformational states that exhibit footprints $a_1$ and
$a_2$, respectively, with $a_1 < a_2$.  We track the position and
state, $(p_j, \sigma_j)$, of each bound protein $j \in [1,N]$. Here,
$p_j \in \{1, 2, \cdots, L\}$ represents the position of the leftmost
nucleotide occupied by protein $j$, as shown in
Fig.~\ref{fig:rpa-schematics-1}(b), \added{and} $\sigma_j \in \{1,
2\}$ labels the state or ``mode'' of protein $j$ (with footprint
$a_{1}$ or $a_{2}$). The exclusion of bound proteins implies the
constraint $p_j + a_{\sigma_j} \leq p_{j+1}$. Adjacent bound proteins
$j$ and $j+1$ are separated by gaps of length $g_j = p_{j+1} - p_j -
a_{\sigma_j} \geq 0$ with the convention that $p_{N+1} = L+1$ and $g_0
= p_1 - 1$. An example of a microstate is shown in
Fig.~\ref{fig:rpa-schematics-1}(b).

For kinetics, we assume that the protein always binds to DNA in the
conformation with a smaller footprint, $a_1$. Then the bound protein
may switch to the conformation with a larger footprint, $a_2$, or
unbind from DNA, as shown in
Fig.~\ref{fig:rpa-schematics-1}(c). Subject to the volume exclusion
effect, the attachment rate $r_{\rm a}$ is given by
\begin{equation}
  r_{\rm a} = k_{\rm a} \sum_{i=0}^N (g_i + 1 - a_1) H(g_i +1 - a_1),
  \label{eq:k_on}
\end{equation}
where $H(x)$ \added{(with $H(0)\equiv 0$)} is the Heaviside step
function. The underlying rationale is that the binding rate is
proportional to the number of possible binding positions. For example,
if $g_i = a_1$, then there is only one possible binding position, and
the binding rate is given by $r_{\rm a} = k_{\rm a}$. If $g_i = a_1 +
1$, then there are two possible binding configurations, and the
binding rate is given by $r_{\rm a} = 2 k_{\rm a}$. For a naked DNA
with only one gap $g_0 = L$, the binding rate is $(L-a_1+1) k_{\rm
  a}$, which remains proportional to system size $L$. We then define
the accessibility of a covered DNA with gap vector $\bm{g}$ to a
protein with footprint $a$ as
\begin{equation}
A(\bm{g} \sep a) = \frac{1}{L} \sum_{i=0}^N (g_i + 1 - a) H(g_i +1 -
a).
\label{eq:accessibility}
\end{equation}
Unlike the overall fraction of vacant lattice sites, $A(\bm{g} \sep
a)$ represents the fraction of lattice positions at which a protein of
footprint $a$ can be placed without overlap.

\deleted{The accessibility decreases exponentially with protein footprint $a$
even when the overall DNA coverage remains unchanged.}
  
Detachment of a bound protein is independent of the gap configuration
$\bm{g}$.  We assume that a bound protein can detach only if it is in
the shorter, more weakly bound state $\sigma_{i} = 1$ so that the
detachment rate is described by
  
\begin{equation}
  r_{\rm d} = k_{\rm d} \sum_{j=1}^N \delta_{\sigma_j, 1} \coloneqq k_{\rm d} N_1,
  \label{eq:k_off}
\end{equation}
where $\delta_{\sigma_j,1}$ is the Kronecker delta function, which is
$1$ if $\sigma_j = 1$ and $0$ otherwise.  Here, $N_1$ is the number of
bound proteins in mode $1$.

A bound protein $j$ in mode $\sigma_j = 1$ can switch to mode $2$ with
rate $k_{12}>0$ only when the gap after the protein \added{$g_j \geq
  (a_2- a_1)$}.  The total $1 \to 2$ transition rate is given by
\begin{equation}
  r_{12} = k_{12} \sum_{j=1}^N \delta_{\sigma_j,1} H\big(g_j + 1 - (a_2 - a_1)\big).
  \label{eq:k_12}
\end{equation}
Proteins in state $\sigma_{j}=2$ convert to state $\sigma_{j}=1$ with
rate $k_{21}$.  As with the detachment of a bound protein in mode
$\sigma_j = 1$, switching from mode $2$ to mode $1$ decreases the
footprint, so this conversion rate is also independent of gap
configuration $\bm{g}$, and is given by
\begin{equation}
  r_{21} = k_{21} \added{\sum_{j=1}^N} \delta_{\sigma_j,2} \coloneqq k_{21} N_2.
  \label{eq:k_off12}
\end{equation}

In addition to attachment and detachment, proteins can also move along
the DNA with rate $D$. Diffusion of a bound protein at position $p_j$
\added{takes} the form of nearest-neighbor jumps, \textit{i.e.}, $ p_j \to p_j
+ 1$ if $g_j \geq 1$ and $ p_j \to p_j - 1$ if $g_{j-1} \geq 1$. The
total rate of diffusion is given by
\begin{equation}
  r_{\rm diff} = D \sum_{j=1}^N \big(H(g_j) + H(g_{j-1})\big).
  \label{eq:k_diff}
\end{equation}

Although one can use the system rates defined in
Eqs.~\eqref{eq:k_on}--\eqref{eq:k_diff} to implement a kinetic
Monte-Carlo algorithm (Gillespie) to simulate RPA binding to ssDNA,
generating stochastic trajectories is computationally intensive for
large systems and fast-diffusing proteins
\cite{Ding2023June}. Therefore, by invoking the fast diffusion limit,
we will formulate a reduced stochastic model which does not require
one to keep track of both the occupancy and gap configurations.

\subsection{Dimension reduction in the fast diffusion limit}

The microscopic rates in Eqs.~\eqref{eq:k_on}--\eqref{eq:k_diff}
depend only on the instantaneous gap configuration and not explicitly
on how \added{the gap statistics} evolve.  Thus, instead of trying to
track every microscopic gap configuration explicitly, we will exploit
the different timescales between rapid diffusion and slower binding,
unbinding, and conformational switching. \added{We first} condition the
instantaneous reaction rates on the numbers of bound proteins in mode
$1$ and $2$, respectively. We find the forms of $r_{\rm d}$ and
$r_{21}$ (Eqs.~\ref{eq:k_off} and \ref{eq:k_off12}) are unchanged, but
the conditioned total binding rate becomes
\begin{equation}
  \begin{aligned}
    r_{\rm a}(N_1, N_2) = &\, k_{\rm a}
    \sum_{\bm{g},\bm{\sigma},i}p(\bm{\sigma},\bm{g}\sep N_1,N_2)   (g_i - a_1 +1)^+
  \\ = &\, k_{\rm a} \sum_{i=0}^{N}\mathbb{E}
  \big[ (g_i-a_1+1)^+ \sep N_1,N_2\big],
  \end{aligned}
  \label{eq:k_on-2}
\end{equation}
where $p(\bm{\sigma},\bm{g}\sep N_1,N_2) $ is the conditional
probability of observing configuration $(\bm{\sigma},\bm{g})$ given
$N_1$ and $N_2$, the summation is over all possible configurations,
and $(g_i - a_1 +1)^+ = \max(g_i - a_1 +1, 0)$ is a condensed way of
writing $(g_i - a_1 +1) H(g_i - a_1 +1)$.


Similarly, the \deleted{conditional} conformation switching rate from mode $1$
to mode $2$ conditioned on $N_1$ and $N_2$ is given by
\begin{equation}
  r_{12}(N_1, N_2) = k_{12}\mathbb{E}\Big[\sum_{j=1}^{N}
 \delta_{\sigma_{j},1}H(g_{j}+1+a_{1}-a_{2})\sep N_{1}, N_{2}\Big].
%
  \label{eq:k_12-2}
\end{equation}
%
%
%

We will take the fast diffusion limit to mean that between successive
binding, unbinding, and conformational switching events, the gap
vector $\bm{g}$ \added{reaches its equilibrium probability
  distribution.}  Since this stationary distribution \added{arises
  from detailed balance of symmetric exclusion dynamics,} it is
uniform over all admissible gap vectors \added{and} is invariant under
permutations of the gap labels. \added{This property allows} the
expectation of sums over gaps to be expressed as $N+1$ times the
expectation over a single arbitrary gap $g_{0}$, \added{which
  simplifies Eqs.~\eqref{eq:k_on-2} and \eqref{eq:k_12-2} to}

\begin{equation}
  \begin{aligned}
  k_{\rm a} \sum_{i=0}^{N}\mathbb{E} & \big[ (g_i-a_1+1)^+ \sep N_1,N_2\big] \\
  \: &  = k_{\rm a} (N+1) \mathbb{E} \big[ (g_0-a_1+1)^+ \sep N_1,N_2\big]
  \end{aligned}
\end{equation}
\added{and}

\begin{equation}
  \begin{aligned}
  k_{12}\mathbb{E}\Big[& \sum_{j=1}^{N}
    \delta_{\sigma_{j},1} H(g_{j}+1+a_{1}-a_{2})\sep N_{1}, N_{2}\Big] \\
  \: & = k_{12} N_1 \mathbb{E} \big[H(g_0 + 1 +a_{1}-a_{2})
    \sep N_1,N_2\big],
  \end{aligned}
  \label{eq:k_12-3}
\end{equation}
\added{respectively.}

To construct a reduced Markov chain model, we explicitly write the
marginal distribution of $g_0$ given $N_1$ and $N_2$.  Upon defining
the number of free sites by $R \equiv L - a_1 N_1 - a_2 N_2$ and the
number of bound proteins by $N = N_1 + N_2$, there are $(N+R)!/(N!
R!)$ total ways of allocating $R$ open sites among $N+1$
gaps. Choosing a representative gap of length $g$ leaves $R-g$ free
sites to distribute among the remaining $N$ gaps in ${R-g+N-1 \choose
  N-1}$ configurations. Since all admissible configurations are
equally likely in equilibrium, we have
\begin{equation}
  \begin{aligned}
  \mathbb{P}(g_0 = g \sep N_1, N_2) & = \frac{{R-g+N-1\choose N-1}}{{N+R
      \choose N}} \\
  \: & = \,\Big(\mfrac{R}{N+R}\Big)^{g}\mfrac{N}{N+R}
  \prod_{j=0}^{g-1}\frac{1-\mfrac{j}{R}}{1-\mfrac{(j+1)}{N+R}}.
  \end{aligned}
  \label{eq:gap-distribution}
\end{equation}
Eq.~\eqref{eq:gap-distribution} is exact for finite system size
\added{with $N \ge 1$. For $N=0$, the single gap has length $L$ and
  $r_{\rm a}=k_{\rm a}(L-a_1+1)^+$.}  \deleted{the last identity
  needed to explicitly specify the reduced stochastic model.}  All
conditional reaction rates can now be computed by taking expectations
over a single gap rather than over the full microscopic configuration
space. \added{We use the combinatorial convention $\binom{u}{v}=0$
  unless $0\le v\le u$, so a transition with insufficient free sites
  has zero rate.} For example, using Eq.~\eqref{eq:gap-distribution}
to explicitly evaluate $\mathbb{E} \big[ (g_0-a_1+1)^+ \sep
  N_1,N_2\big]$ appearing in $r_{\rm a}(N_1, N_2)$, \added{we find}
\begin{equation}
  \begin{aligned}
   \mathbb{E}\big[ (g_0-a_1 & +1)^+ \sep N_1,N_2 \big]
   \\ \: & = \sum_{g=0}^{R}(g-a_{1}+1)^{+}
     \mathbb{P}\big[g_{0}=g\sep N_{1}, N_{2}\big],
  \end{aligned}
\end{equation}
which gives us an explicit expression for the attachment rate
conditioned on the state variables $N_{1}, N_{2}$:

\begin{equation}
  \begin{aligned}
r_{\rm a}(N_{1}, N_{2}) =
    k_{\rm a}(N+1)\sum_{g=0}^{R}(g-a_{1}+1)^{+}\frac{{R-g+N-1\choose
        N-1}}{{N+R \choose N}}.
  \end{aligned}
  \label{eq:r1_explicit}
\end{equation}
\added{Eq.~\ref{eq:r1_explicit} can be further manipulated into
  the convenient form}

\begin{equation}
  \begin{aligned}
\added{r_{\rm a}(N_{1}, N_{2}) = k_{\rm a}(N+1) \frac{{N+R+1-a_{1}\choose
    N+1}}{{N+R\choose N}}.}
  \end{aligned}
  \label{eq:r1_explicit2}
\end{equation}

Using Eq.~\eqref{eq:gap-distribution} and
\begin{equation}
  \begin{aligned}
\mathbb{E} \big[H(g_0 + 1 + & a_{1}-a_{2}) 
  \sep N_1,N_2\big] \\
\equiv & \,\mathbb{P}(g_0 > a_2 - a_{1}-1 \sep N_1, N_2) \\
= & \, \added{\mfrac{{N+R-(a_{2}-a_{1}) \choose N}}{{N+R \choose N}}}
  \end{aligned}
  \label{eq:gap-tail}
\end{equation}
\added{in Eqs.~\eqref{eq:k_12-3} and \eqref{eq:k_12-2}, the
  conditional gap-averaged conformational switching rate $r_{12}(N_1,
  N_2)$ reduces to}
\begin{equation}
 r_{12}(N_{1}, N_{2}) = k_{12}N_{1}\frac{{N+R-(a_{2}-a_{1})\choose
     \added{N}}}{{N+R\choose N}}.
  \label{eq:r12_explicit}
\end{equation}
The conditional rates in Eqs.~\eqref{eq:r1_explicit},
\eqref{eq:r12_explicit}, and the unconstrained rates in
Eqs.~\eqref{eq:k_off} and \eqref{eq:k_off12} allow us to define an
effective Markov model over the state space ($N_{1}, N_{2}$) that is
valid in the fast diffusion limit.

\subsection{Hydrodynamic limit}
Eq.~\eqref{eq:gap-distribution} is \added{cumbersome to evaluate for
  large} $L$, so we take the hydrodynamic limit in order to obtain a
closed form expression for $r_{\rm a}(N_1, N_2)$ and $r_{12}(N_1,
N_2)$. For large $L$, we rescale the variables $\rho_1\equiv N_1/L$,
$\rho_2\equiv N_2/L$, define the total particle number density $\rho
\equiv \rho_1 + \rho_2$ and the \added{fractional site occupancy
  (``coverage'') $\theta \equiv a_{1}\rho_{1}+a_{2}\rho_{2} =
  (a_{1}N_{1}+a_{2}N_{2})/L$ to find}

\begin{equation}
  \mathbb{P}(g_0 = g \sep \rho_1, \rho_2) = 
  \added{(1-\eta)\eta^{g}  \prod_{j=0}^{g-1}
    \frac{1-\mfrac{j}{R}}{1-\mfrac{(j+1)}{N+R}},}
%
  \label{eq:gap-distribution-hydro0}
\end{equation}
\added{where}

\begin{equation}
  \begin{aligned}
    \added{\eta \equiv \mfrac{R}{N+R}} &  \added{ =
      \mfrac{L-a_{1}N_{1}-a_{2}N_{2}}{L-(a_{1}-1)N_{1}-(a_{2}-1)N_{2}}} \\
  \: & \added{ =
    \mfrac{1-a_{1}\rho_{1}-a_{2}\rho_{2}}{1-(a_{1}-1)\rho_{1}-(a_{2}-1)\rho_{2}}} \\
  \: &  \added{= \mfrac{1-\theta}{1-\theta+\rho}. }
  \end{aligned}
  \label{eq:eta}
\end{equation}
\added{The last product term in Eq.~\ref{eq:gap-distribution-hydro0} is
approximately unity for $g^{2} \ll R(N+R)/N$, resulting in}

\begin{equation}
  \mathbb{P}(g_0 = g \sep \rho_1, \rho_2) \approx
  (1-\eta)\eta^{g} \equiv \frac{\rho(1-\theta)^{g}}{(1-\theta+\rho)^{g+1}}.
  \label{eq:gap-distribution-hydro}
\end{equation}
The appearance of a geometric distribution reflects the absence of
long-range spatial correlations in the fast-diffusion
limit. \replaced{An intuitive picture of the geometric gap
  distribution comes from mapping the rods onto unit-length particles
  on a ``compressed'' lattice. At fixed protein number and ordered
  modes, this mapping preserves all gap lengths and gives an ideal
  lattice gas on $R+N$ sites. The fixed particle number correlates
  site occupancies, but these correlations weaken as the system grows
  at fixed density.  Treating successive compressed-lattice sites as
  approximately independent, with vacancy probability $\eta=R/(R+N)$,
  yields the geometric gap distribution $(1-\eta)\eta^g$.}{Since
  diffusion randomizes protein positions, successive empty sites occur
  independently on sufficiently large scales and make the gap lengths
  geometrically distributed in the thermodynamic limit.} For
sufficiently large $L$, we find the following asymptotic expressions
for $r_{\rm a}(N_1, N_2)$ and $r_{12}(N_1, N_2)$:
\begin{equation}
  \begin{aligned}
    r_{\rm a}(N_1, N_2) &\sim k_{\rm a} \eta^{a_1-1} (1-\theta)L =
    k_{\rm a} \eta^{a_1-1} R,\\
    r_{12}(N_1, N_2) & \sim k_{12} \eta^{a_2-a_1} N_1.
  \end{aligned}
  \label{eq:hydrodynamic}
\end{equation}
A protein with footprint $a$ can bind only \added{if $a-1$ additional}
consecutive sites \added{(beyond the first vacant site)} are
vacant. Under the geometric gap distribution, each additional required
empty site contributes an independent factor of $\eta$, giving the
exponential suppression of accessibility with increasing footprint.
This size-dependent selection against \added{larger footprints,
  captured by the factor $\eta^{a-1}$, was recognized by} McGhee and
von Hippel \cite{McGhee1974June}. The mean accessibility \added{for}
an object of size $a$ is found by taking the expectation of $A(\bm{g}
\sep a)$ (Eq.~\ref{eq:accessibility}) with respect to the distribution
$\mathbb{P}(g_0 = g \sep \rho_1, \rho_2)$
(Eq.~\ref{eq:gap-distribution-hydro}). From permutation symmetry
(identically distributed $g_{i}$), we find

\begin{equation}
  \begin{aligned}
\mathbb{E}[A(\rho,\theta \sep a)] \approx  \frac{N+1}{L}&
    \sum_{g=a}^{\infty}(g-a+1)(1-\eta)\eta^{g} \\
    \xrightarrow[L \to \infty]{} & \, \frac{\rho\eta^{a}}{1-\eta} =
    \frac{(1-\theta)^{a}}{(1-\theta + \rho)^{a-1}}.
\end{aligned}
  \label{eq:A2}
\end{equation}
Written directly in terms of the system parameters, the mean
accessibility is $\mathbb{E}[A(\rho_1, \rho_2, a_1, a_2 \sep a)]$,
where $a$ is the size of a particle attempting to attach.  Since our
model assumes attachment occurs only in mode 1, we set $a=a_{1}$.  The
quantity $\mathbb{E}[A(\rho_1, \rho_2, a_1, a_2 \sep a_{1})]$
incorporates the effect of volume exclusion on the total attachment
rate.

Upon combining Eqs.~\eqref{eq:hydrodynamic}, \eqref{eq:k_off}, and
\eqref{eq:k_off12}, we derive a complete analytical description of the
continuous-time Markov chain model for the reduced system $\left\{
(N_1, N_2) \in \mathbb{N}^2: a_1N_1 + a_2N_2 \leq L \right\}$ when a
sufficiently large but finite $L$ is used.  Specifically, we arrive at
the following discrete reduced-state master equation
%
%
%
\begin{widetext}
\begin{equation}
  \begin{aligned}
    \frac{\partial P(N_1,N_2,t)}{\partial t} = & \added{r_{\rm a}(N_{1}-1, N_{2})}
    P(N_1-1,N_2,t) + \added{r_{12}(N_{1}+1, N_{2}-1)}P(N_1+1,N_2-1,t)\\
    \: & + \added{r_{\rm d}(N_1+1)} P(N_1+1,N_2,t) +
    \added{r_{21}(N_2+1)} P(N_1-1,N_2+1,t)\\
    \: & -\added{\big[r_{\rm a}(N_{1}, N_{2})+ r_{12}(N_{1}, N_{2})
        +r_{\rm d}(N_{1})+r_{21}(N_{2})\big]} P(N_1,N_2,t).
\end{aligned}
\label{eq:master-equation}
\end{equation}
\end{widetext}
\added{We set transition rates into inadmissible states to
  zero. Starting from inside the state space $\mathcal{S}_{L} =
  \left\{ N_1, N_2 \geq 0\;\middle\vert{}\; a_1 N_1 + a_2 N_2 \le L
  \right\}$, $P(N_{1}, N_{2}, t)=0$ for all $(N_{1}, N_{2}) \notin
  \mathcal{S}_{L}$.}

\added{The rates $r_{\rm d}(N_{1}) \equiv k_{\rm d}N_{1}$ and
  $r_{21}(N_{2})\equiv k_{21}N_{2}$ are developed in
  Eqs.~\eqref{eq:k_off} and ~\eqref{eq:k_off12}, respectively. One can
  use Eqs.~\eqref{eq:r1_explicit2} and \eqref{eq:r12_explicit} for the
  attachment and mode 1 $\to$ mode 2 transition rates $r_{\rm
    a}(N_{1}, N_{2})$ and $r_{12}(N_{1}, N_{2})$,
  respectively. Alternatively the hydrodynamic approximations in
  Eq.~\eqref{eq:hydrodynamic} can be used. Since $\eta$ is given by
  Eq.~\ref{eq:eta} and $R\equiv L - a_{1}N_{1}-a_{2}N_{2}$, the
  discrete master equation \eqref{eq:master-equation} for a Markov
  process within the $(N_1, N_2)$ state space can still be defined in
  the hydrodynamic, large $L$ limit. Using
  Eqs.~\eqref{eq:hydrodynamic} in Eq.~\eqref{eq:master-equation} and
  applying a Kramers-Moyal expansion allows a direct path to the large
  $N_{1}, N_{2}$ SDEs for $\rho_{1}$ and $\rho_{2}$:}

\begin{equation}
  \begin{aligned}
\added{\mathrm{d}\rho_1 =} & \added{\left[
k_{\rm a}(1-\theta)\eta^{a_1-1}
-k_{\rm d}\rho_1 -k_{12}\eta^{a_2-a_1}\rho_1+k_{21}\rho_2\right]\mathrm{d} t} \\
\: & \quad \added{+ \frac{1}{\sqrt{L}}\sqrt{k_{\rm a}(1-\theta)\eta^{a_1-1}
+k_{\rm d}\rho_1}\,\mathrm{d} W_1} \\
\: &\quad  \added{-\frac{1}{\sqrt{L}}
\sqrt{k_{12}\rho_1\eta^{a_2-a_1}+k_{21}\rho_2}\,\mathrm{d} W_2,} \\
\added{\mathrm{d} \rho_2 =} & \added{\left[k_{12}\rho_1\eta^{a_2-a_1}-k_{21}\rho_2
\right] \mathrm{d} t} \\
\: & \quad \added{+ \frac{1}{\sqrt{L}}
\sqrt{k_{12}\rho_1\eta^{a_2-a_1}+k_{21}\rho_2}\,\mathrm{d} W_2,}
\end{aligned}
\label{eq:hydro_SDE}
\end{equation}
\added{where $\mathrm{d} W_{1}$ and $\mathrm{d} W_{2}$ are independent
  Wiener processes, and $\eta =
  \mfrac{1-a_{1}\rho_{1}-a_{2}\rho_{2}}{1-(a_{1}-1)\rho_{1}-(a_{2}-1)\rho_{2}}$
  (from Eq.~\ref{eq:eta}). Since Eq.~\eqref{eq:hydro_SDE} is not
  accurate when $\rho_{i} \to 0$ and $\rho_{1}a_{1}+\rho_{2}a_{2} \to
  1$, additional conditions are required to establish reflecting
  boundary conditions at $\rho_{1}=0, \rho_{2}=0$ and
  $\rho_{1}a_{1}+\rho_{2}a_{2} = 1$ for the associated Fokker-Planck
  equation \cite{lions1984}.}

\added{In the $L\to \infty$ limit, the distribution functions sharpen
  by the law of large numbers. Neglecting the noise terms in
  Eqs.~\eqref{eq:hydro_SDE}, we find the deterministic evolution of
  $\rho_{i}(t)$ follows the coupled ODEs}
%
  \begin{equation}
  \begin{aligned}
    \frac{\mathrm{d} \rho_1}{\mathrm{d} t} & =
    \added{\frac{k_{\rm a}\big(1-a_{1}\rho_{1}-a_{2}\rho_{2}\big)^{a_{1}}}{\big(1-(a_{1}-1)\rho_{1}
-(a_{2}-1)\rho_{2}\big)^{a_{1}-1}} } -
    k_{\rm d} \rho_1  \\
    \: & \,\quad - k_{12}
    \added{
      \left(\mfrac{1-a_{1}\rho_{1}-a_{2}\rho_{2}}{1-(a_{1}-1)\rho_{1}-(a_{2}-1)\rho_{2}}
    \right)^{a_2-a_1}  }
\!\!\!\! \rho_1 + k_{21} \rho_2,
\\
\frac{\mathrm{d} \rho_2}{\mathrm{d} t} & = k_{12}
\added{\left(\mfrac{1-a_{1}\rho_{1}-a_{2}\rho_{2}}
     {1-(a_{1}-1)\rho_{1}-(a_{2}-1)\rho_{2}}\right)^{a_2-a_1}} \!\!\!\rho_{1}
     - k_{21} \rho_2.
  \end{aligned}
  \label{eq:hydrodynamic-ode}
\end{equation}
%
\deleted{associated with the infinite-$L$ asymptotic limit of the
  reduced-state Markov model (Eq.~\ref{eq:master-equation}).  Note
  that Eqs.~\eqref{eq:hydrodynamic-ode} is a closed set of nonlinear
  ODEs for $\rho_{1}(t), \rho_{2}(t)$ since $\theta \equiv
  \rho_{1}a_{1}+\rho_{2}a_{2}$.} Eqs.~\eqref{eq:master-equation},
\eqref{eq:hydro_SDE}, and \eqref{eq:hydrodynamic-ode} are appropriate
models only in the infinitely fast diffusion limit and are thus
independent of $D$. \added{We have verified that provided $N_{1},
  N_{2} \gg 1$, the SDE/Fokker-Planck model yields predictions with
  very accurate mean and variance relative to those generated by
  Gillespie Monte-Carlo simulations of the full-lattice model with
  fast diffusion.  Therefore, we will analyze results only from
  full-lattice Gillespie Monte-Carlo simulations with diffusion, the
  large-$D$ limit master equation~\eqref{eq:master-equation}, and the
  hydrodynamic model Eq.~\eqref{eq:hydrodynamic-ode}.}

\subsection{Fast diffusion requirement}
\label{sec:fast-diffusion-requirement}

In our derivation of Eqs.~\eqref{eq:master-equation} and
\eqref{eq:hydrodynamic-ode}, we have assumed that the diffusion of RPA
is fast enough that \replaced{positions conditionally equilibrate
  between reactions, giving the finite-size gap law in
  Eq.~\eqref{eq:gap-distribution} and its large-system geometric limit
  in Eq.~\eqref{eq:gap-distribution-hydro}.}{the system is well-mixed
  and the gap size distribution is geometric} Since diffusivity $D$
and reaction rates describe different processes (have different
units), we motivate here more rigorous criteria under which local
equilibrium is established sufficiently rapidly so that our reduced
models remain accurate.

One issue is that the rate of convergence decreases with the
DNA length $L$. Consider diffusion of a single particle with a
footprint $a$ on a lattice with a finite length $L$. Its diffusion
along the lattice is identical to that of a particle of unit
footprint moving on a lattice with length $L-(a-1)$. The transition
matrix $M$ can be explicitly written as

\[
\bm{M} = D \begin{pmatrix}
    -1 & 1 & 0 & \cdots & \cdots & 0 \\
    1 & -2 & 1 & \ddots & & \vdots \\
    0 & 1 & -2 & \ddots & \ddots & \vdots \\
    \vdots & \ddots & \ddots & \ddots & 1 & 0 \\
    \vdots & & \ddots & 1 & -2 & 1 \\
    0 & \cdots & \cdots & 0 & 1 & -1
    \end{pmatrix},
\]
where $D$ is the diffusivity. The eigenvalues of $\bm{M}$ are given by
$\lambda_k = -2D \big[ 1 - \cos(k\pi/(L-a+1)) \big]$, where $\lambda_1
= -2D \big[ 1 - \cos(\pi/(L-a+1)) \big]$ is the largest nonzero
eigenvalue. Since $\lambda_1 \sim -D\pi^2/L^2$ as $L \rightarrow
\infty$, the rate of convergence decreases quadratically with
  $L$. A similar argument can be applied to $N$ particles, each of
  footprint $a$, on a lattice of length $L$.

According to Aldous' spectral gap conjecture, proved in
Ref.~[\onlinecite{Pietro2009Jun}], the \replaced{positive relaxation rate (spectral
  gap)}{largest nonzero eigenvalue (spectral gap)} of this
indistinguishable-particle system is
\replaced{$4D\sin^2\!\big[\pi/(2[L-N(a-1)])\big]$}{$-2D\big[1-\cos(\pi/(L-Na+1))\big]$},
identical to that of a single-particle system \replaced{on the
  compressed chain of length $L-N(a-1)$}{in which the particle has
  length equal to the combined length $Na$ of all $N$ particles}.
\added{This comparison applies when the compressed chain contains both
  particles and vacancies.} As a result, diffusion can never be fast
enough to achieve equilibration across the entire, arbitrarily long
system.  However, the relevant question is whether the local region
near a kinetic event can approximately reach equilibrium,
\replaced{which motivates a local-equilibration criterion}{Such local
  thermodynamic equilibrium is sufficient for the fast diffusion
  approximation} because binding and conformational changes depend
only on nearby gap configurations rather than on the global
arrangements of proteins and gaps.

\deleted{It was shown by de Gier and Essler that for large system size
  $L$, symmetric simple exclusion processes with unit footprint and
  open boundaries have a spectral gap of the form $\lambda_1 \sim
  -D\pi^2/L^2$ \cite{Gier2006}. We then conjecture that a similar
  spectral gap holds for systems with larger footprints. It is then
  possible to consider local equilibria by separating a long chain
  into disjoint sections of length $\Delta$.  The additional
  constraint $\Delta \gg a$ ensures that most kinetic events occur
  within the section rather than spanning adjacent sections. We thus
  also need the relaxation rate within each section $\Delta$ to be
  much faster than any kinetic rate $k^{+}={\rm max}\{k_{\rm a},
  k_{12}\}$ and $k^{-}={\rm max}\{k_{\rm d}, k_{21}\}$. This condition
  can be expressed as $D \pi^2/\Delta^2 \gg \max \{k^{+} \Delta, k^{-}
  (\Delta/a) \}$ such that the local system can reach equilibrium
  before the next kinetic event. A sufficient condition for the prior
  fast diffusion limit approximation to be valid is $D \gg \max
  \{k^{+} a^3, k^{-} a^2\}$.}

\added{To extend this comparison to both protein footprints, consider
  a section of length $L'$ with reflecting ends, a fixed protein
  number $N'$, and fixed left-to-right mode sequence
  $\boldsymbol{\sigma}$. Diffusion changes positions within this
  section while preserving these quantities. If $N_i'$ are the $i^{\rm
    th}$-mode bound protein counts within the section, we can map the
  hard-rod positions onto those of unit-length particles on a closed
  section of length $L''=L'-(a_1-1)N_1'-(a_2-1)N_2'=R'+N'$ where $R'
  \equiv L'-a_1N_1'-a_2N_2'$.}
%
%
\added{The same spectral-gap identity applies to this compressed
  chain.  Defining the quantities $\rho_i'=N_i'/L'$,
  $\rho'=\rho_1'+\rho_2'$, and $\theta'=a_1\rho_1'+a_2\rho_2'$ within the
  section, the diffusion-only spectral gap is}
\begin{equation}
\added{\lambda' = 4D\sin^2\!\left(\frac{\pi}{2L''}\right)
\simeq D\Big(\frac{\pi}{L''}\Big)^{2},}
\label{eq:local-diffusion-gap}
\end{equation}
\added{where the approximation requires $L'' \gg 1$. Sections with
  $N'=0$ or $N'=L''$ have a unique gap arrangement.}

\added{Define the accessibility and switching availability of each section by}
\[
  \begin{aligned}
A' & =\mfrac{1}{L'}\sum_{j=0}^{N'}(g_j+1-a_1)H(g_j+1-a_1),\\
S' & =\sum_{j=1}^{N'}\delta_{\sigma_j,1}H(g_j+1-a_2+a_1).
  \end{aligned}
  \]
\added{Here, $g_0,\ldots,g_{N'}$ are the ordered gaps with $g_0=L'$
  for an empty section ($N'=0$).  Then, $r_{\rm a}=k_{\rm a} L' A'$,
  $r_{\rm d}=k_{\rm d}N_1'$, $r_{12}=k_{12}S'$, and
  $r_{21}=k_{21}N_2'$. The total rate $r_{\rm tot}$ can be shown to
  satisfy $r_{\rm tot}\leq L' K(\rho_1',\rho_2')$ where $K\equiv
  k_{\rm a}(1-\theta')+(k_{\rm d}+k_{12})\rho_1'+k_{21}\rho_2'$.  This
  bound holds regardless of the gap configuration since it assumes
  spacings of proteins within a section that maximizes the number of
  allowed transitions. If we use equilibrium-averaged spacings
  instead, replacing $A'$ and $S'$ by their equilibrium averages, a
  less conservative, but more complex, coverage-dependent condition
  could be derived. Comparing the simple stringent bound on the total
  intra-section reaction rate with the relaxation rate $\lambda'$ in
  Eq.~\eqref{eq:local-diffusion-gap} defines $D \gg L'
  \big(L''/\pi\big)^{2} K(\rho_{1}', \rho_{2}')$ as a fast diffusion
  condition.  The nonnegative weights $1-\theta'$, $a_1\rho_1'$ and
  $a_2\rho_2'$ sum to one so $K \le K_* \equiv \max\{k_{\rm a},(k_{\rm
    d}+k_{12})/a_1,k_{21}/a_2\}$. Using $L''/L' \le 1$ and setting
  $a_* \equiv \max\{a_1,a_2-a_1\}$ as the largest length that
  attachment or extension requires, we can define $N_{*}' \equiv
  L'/a_*$ as the number of $a_{*}$ lengths in the section. The
  density-independent fast diffusion criterion becomes}

\begin{equation}
\added{D\gg\frac{(L')^{3}}{\pi^2}
\max\left\{k_{\rm a},\frac{k_{\rm d}+k_{12}}{a_1},\frac{k_{21}}{a_2}\right\}.}
\label{eq:fast-diffusion-envelope}
\end{equation}
\added{The separate bounds on $L''/L'$ and $K$ make this estimate
  conservative. It is a heuristic local-equilibration criterion in
  which $N_{*}'=1$ represents only the smallest section lengthscale
  $L'$ that allows kinetic transitions. Strong kinetic timescale
  separation, in which sections are long enough to allow many kinetic
  steps, but not so large that diffusion is unable to equilibrate
  them, requires $1 \ll N_{*}' \ll[\pi^2D/(a_*^3K_*)]^{1/3}$. For the
  footprints used in our comparisons, $a_1\ge a_2-a_1$, so $a_*=a_1$.
  Choosing $N_{*}'=1$ and replacing strong scale separation by a
  heuristic comparison delimits the crossover to fast diffusion
  according to a cubic/quadratic form}

\begin{equation}
\added{D \ge \mfrac{1}{\pi^2}\max\left\{
k_{\rm a}a_1^3,\;(k_{\rm d}+k_{12})a_1^2,\;
k_{21}\frac{a_1^3}{a_2}\right\}.}
\label{eq:practical-diffusion-region}
\end{equation}
\added{The attachment term requires $D \ge k_{\rm a}a_1^3/\pi^2$ while
  the other processes require $k_{12} < \pi^2D/a_1^2-k_{\rm d}$ and
  $k_{21}<\pi^2Da_2/a_1^3$. These boundaries for $k_{12}$ and $k_{21}$
  are indicated by the dashed lines in Fig.~\ref{fig:wasserstein}.}

\subsection{\added{Equilibrium with and without diffusion}}
\label{sec:equilibrium-without-diffusion}

\added{Detachment followed by reattachment elsewhere can relocate an
  indistinguishable protein, leading to the same equilibrium state
  independent of diffusion along the lattice. Let $\Omega$ be the
  finite space of admissible configurations and write
  $\omega=(\boldsymbol{\sigma},\boldsymbol{g})\in\Omega$.  With
  $N_i(\omega)=\sum_{j=1}^{N(\omega)}\mathbf{1}_{\{\sigma_j=i\}},
  \quad i=1,2$, the counts of proteins in mode $i$, the equilibrium
  distribution can be written as}

\begin{equation}
 \added{ \pi(\omega)=\frac{z_1^{N_1(\omega)}z_2^{N_2(\omega)}}{Z},
   \quad z_1\equiv \mfrac{k_{\rm a}}{k_{\rm d}}, \,\,\, z_2\equiv
   z_1\mfrac{k_{12}}{k_{21}}.}
\end{equation}
\added{Here, $N(\omega)=N_1(\omega)+N_2(\omega)$ and
  $Z=\sum_{\omega\in\Omega}z_1^{N_1(\omega)}z_2^{N_2(\omega)}$ is the
  partition function.  Denote configurations reached by an allowed
  attachment, expansion, or hop by $\omega_{\rm a}$, $\omega_{12}$,
  and $\omega_{\rm h}$, respectively. Their weights satisfy}

\begin{equation}
\begin{aligned}
\added{\pi(\omega)k_{\rm a}}&=\added{\pi(\omega_{\rm a})k_{\rm d}},\\
\added{\pi(\omega)k_{12}}&=\added{\pi(\omega_{12})k_{21}},\\
\added{\pi(\omega)D}&=\added{\pi(\omega_{\rm h})D}.
\end{aligned}
\label{eq:equilibrium-any-D}
\end{equation}
\added{Each transition type therefore preserves the same distribution.
  With positive reaction rates, contraction and detachment connect all
  configurations to the empty lattice, giving a unique equilibrium
  even when $D=0$. Its weights are uniform over admissible positions
  conditional on protein number and ordered modes.}

\section{Results}

We compare predictions of our effective models
(Eqs.~\ref{eq:master-equation} and \ref{eq:hydrodynamic-ode}) against
each other and against \replaced{reference lattice
  simulations}{``ground-truth'' simulations} to show the accuracy of
our approximations. Then, we refit existing experimental data to
illustrate the improvement our ODE model exhibits over previous
models. Finally, we generalize our model to proteins that unbind by
peeling off incrementally and bind by successively forming contact
bonds. This type of association/dissociation dynamics extends our
analysis to include a wider class of DNA-binding proteins such as
histones.

\subsection{Model validation}
We numerically implemented and simulated the full and reduced
stochastic models, as well as the deterministic hydrodynamic
models. To compare the models and test whether the fast diffusion
limit is valid, we quantify the differences between the predictions of
each model by averaging the squared Wasserstein distance between the
empirical distributions of sample trajectories at each timepoint
during the simulation.
\begin{figure}[htb]
  \centering
  \includegraphics[width=3.0in]{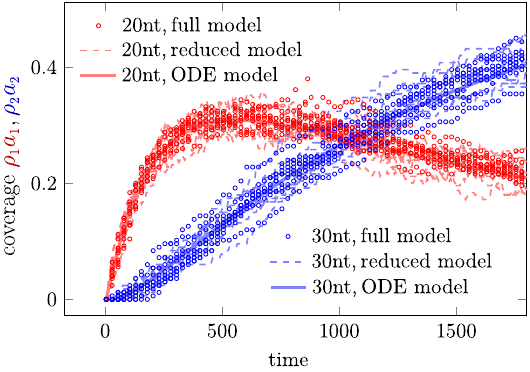}
  \caption{Comparisons of the occupation fractions $\rho_{1}a_{1},
    \rho_{2}a_{2}$ predicted by the full-configuration space
    agent-based model (simulation points), the reduced-state Markov
    model (dashed curves), and the deterministic hydrodynamic model
    (solid curves). The parameters used are $k_{\rm a} = 10^{-4}$/s,
    $k_{\rm d} = 10^{-3}$/s, $k_{12}=10^{-3}$/s, $k_{21}=10^{-4}$/s,
    \replaced{$L=5000$ nt}{$L=500$ nt}, $a_1=20$, $a_2=30$, and
    \replaced{$D=0.1$}{$D=0.01$} \added{nt}$^2$/s. The squared
    Wasserstein-2 distance between the reduced-state Markov chain
    model and the full agent-based model is \added{$3.25 \times
      10^{-5}$, while the squared Wasserstein-2 distance between the
      full agent-based model and the deterministic ODE model is $1.4
      \times 10^{-3}$.}}
  \label{fig:trajectories}
\end{figure}

As an illustrative example, we choose a \added{representative
  parameter set} and plot the \added{corresponding} trajectories of
the coverages $\rho_{1}a_{1}, \rho_{2}a_{2}$ from the three models in
Fig.~\ref{fig:trajectories}. \added{This parameter set satisfies the
  practical fast-diffusion criterion in
  Eq.~\eqref{eq:practical-diffusion-region}, with $D=0.1>0.0811$
  nt$^2$/s at $N_{*}'=1$.} As shown in the Supplementary Information
(SI), all three models exhibit similar coverage dynamics, but the
reduced-state Markov chain model maintains \added{a level of variance
  similar to that of} the full agent-based model at each timepoint;
the reduced-state Markov model has a smaller squared Wasserstein
distance to the full agent-based model than does the deterministic
hydrodynamic model. \added{For this parameter set,
  Fig.~\ref{fig:supp-mean-variance} in the SI separately plots
  the mean and variance of the coverage trajectories. The three models
  have closely matching means, while the reduced Markov model also
  captures the nonzero coverage variances of the full model.}

\begin{figure}[!t]
  \centering
  \includegraphics[width=3.35in]{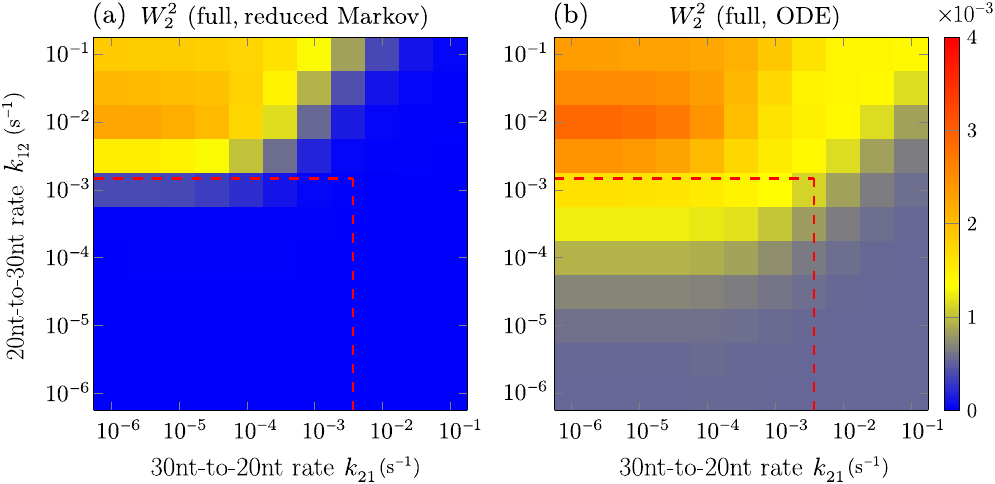}
  \caption{Heatmap of the squared Wasserstein-2 distance between (a)
    the reduced Markov chain model and the fully simulated agent-based
    model, and (b) the deterministic hydrodynamic ODE model and the
    fully simulated agent-based model, as a function of $k_{12}$ and
    $k_{21}$. The parameters $k_{\rm a} = 10^{-4}$/s, $k_{\rm d} =
    10^{-3}$/s, \added{$L=5000$ nt,} $a_1=20$, $a_2=30$, and
    \replaced{$D=0.1$}{$D=0.01$} \added{nt}$^2$/s are fixed. The
    squared Wasserstein-2 \added{distance uses squared Euclidean cost
      in the fractional-coverage space $(a_1\rho_1,a_2\rho_2)$. Each
      stochastic ensemble contains 4000 trajectories and at each
      ($k_{12}, k_{21}$) the squared Wasserstein-2 distance between two
      ensembles of 4000 values each is computed. These values are
      obtained at each timepoint. The plotted values are obtained by
      averaging over the 360 timepoints $t=0,5, \ldots, 1795$ s.}
    \added{The dashed red boundaries represent the $N_{*}'=1$ crossover
      from the density-independent local-equilibration estimate.}}
  \label{fig:wasserstein}
\end{figure}
\begin{figure*}[!t]
  \centering
  \includegraphics[width=5.8in]{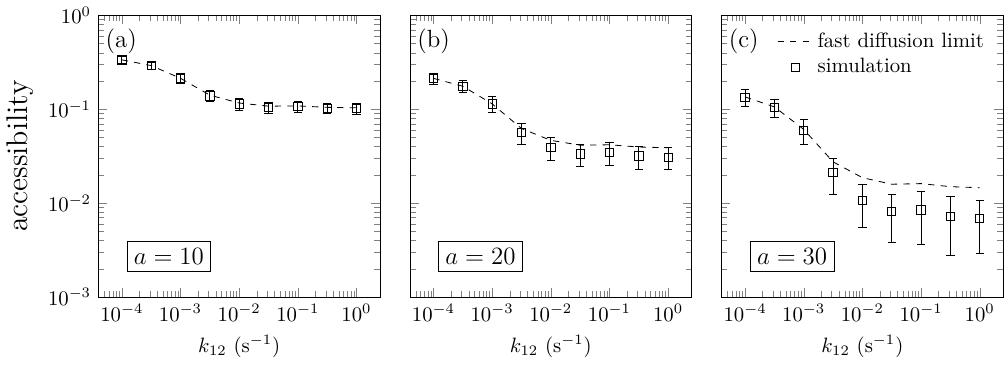}
  \caption{In the presence of two-mode ($a_{1}=20$ and $a_{2}=30$) RPA
    molecules, the accessibility of ssDNA to additional DNA-binding
    proteins of footprints 10 (a), 20 (b), and 30 (c), respectively,
    plotted as functions of $k_{12}$.  Here, $k_{\rm a} = 10^{-4}$/s,
    $k_{\rm d} = 10^{-3}$/s, $D=0.01$ \added{nt}$^2$/s, and
    $k_{21}=10^{-6}$/s. The results are obtained from 100 sample
    trajectories for each protein of size $a$ at fixed time $t=1800$
    s. After obtaining $\theta$ and $\rho$ from simulation
    trajectories, we \replaced{evaluated the hydrodynamic
      accessibility estimate at their ensemble means}{used them to
      compute the mean accessibility} $\mathbb{E}[A(\rho, \theta \sep
      a)]=(1-\theta)^{a}/(1-\theta+\rho)^{a-1}$ (Eq.~\ref{eq:A2}) in
    the fast diffusion limit, as shown by the dashed curves. The
    actual accessibility of a configuration was evaluated by counting
    the total number of positions accessible by an object of footprint
    $a$. After normalizing by $L=5000$ nt, we plot their means (open
    square symbols) and standard deviations (error bars) as a function
    of $k_{12}$, alongside $\mathbb{E}[A(\rho,\theta \sep
      a)]=(1-\theta)^{a}/(1-\theta+\rho)^{a-1}$ from
    Eq.~\eqref{eq:A2}.}
  \label{fig:gap_accessibility}
\end{figure*}
In Fig.~\ref{fig:wasserstein}, \added{we compare trajectories of
  $\boldsymbol\rho(t) = \big(\rho_{1}(t), \rho_{2}(t)\big)$ predicted
  by the reduced Markov and ODE models to those derived from
  simulations of the agent-based model as $k_{12}$ and $k_{21}$ are
  varied. The parameters} $k_{\rm a} = 10^{-4}$/s, $k_{\rm d} =
10^{-3}$/s, \replaced{$L=5000$nt}{$L=500$nt}, $a_1=20$, $a_2=30$, and
$D=0.1$ \added{nt}$^2$/s are fixed. At each sampled time, we formed
the empirical distribution of the \replaced{fractional-coverage vector
  $(a_1\rho_1,a_2\rho_2)$ from 4000 independently simulated
  trajectories}{two-dimensional number density vector
  $\boldsymbol\rho(t)$ from 100 independently simulated trajectories}.
We computed the squared Wasserstein-2 distance between the empirical
distributions generated by the \added{full agent-based} and reduced
Markov models. \added{The values reported in both panels of
  Fig.~\ref{fig:wasserstein} are the arithmetic mean of these snapshot
  distances over all sampled times.} \added{The red dashed lines
  correspond to the crossover to fast diffusion defined by $N_{*}'=1$
  from the revised density-independent estimate
  (Eq.~\ref{eq:practical-diffusion-region}).  The reduced Markov model
  retains noise (which the deterministic ODE model neglects) and
  generates coverages that agree well with those from the full-model
  simulation in the indicated parameter regime.}

\paragraph{\added{Reaction-mediated redistribution.}}
\label{sec:reaction-redistribution}
\added{Fig.~\ref{fig:wasserstein} also shows good agreement outside
  the region indicated by the diffusion-based criterion. This
  agreement can arise because detachment followed by attachment
  elsewhere relocates an indistinguishable protein and changes the
  neighboring gaps. The detailed-balance argument in
  Eq.~\eqref{eq:equilibrium-any-D} shows that the full model has the
  same unique equilibrium for every $D$, including $D=0$.  This common
  equilibrium allows agreement without rapid lateral diffusion,
  although the time needed to approach it also depends on the reaction
  rates.}

\added{Faster detachment makes this redistribution pathway more
  accessible. As shown by Fig.~\ref{fig:supp-unbinding} in the
  Supplementary Information (SI), the coverage trajectories with and
  without diffusion become closer as $k_{\rm d}$ increases.  When
  detachment is slow, the diffusionless system overshoots its
  equilibrium coverage before relaxing, reflecting a transient excess
  of mode-1 proteins. A protein in mode 2 must first convert to mode 1
  at rate $k_{21}$ before it can detach, so slow conversion can also
  delay relaxation. Fig.~\ref{fig:supp-contraction} compares
  trajectories with and without diffusion at different values of
  $k_{21}$. Faster conversion brings the trajectories closer, whereas
  slower conversion produces persistent transient differences.
  Fig.~\ref{fig:supp-binding} shows that faster detachment generally
  improves agreement between distributions predicted by different
  models. With slow detachment and rapid attachment, proteins instead
  accumulate before substantial rearrangement occurs, preserving the
  gap biases of sequential adsorption. These reaction pathways explain
  why the diffusion-based crossover can miss parameter sets with good
  finite-time agreement.}

When diffusion does not compensate for local disruption caused by
kinetic events, the accessibility of DNA to larger proteins \added{is
  reduced more, relative to the fast-diffusion prediction, than that
  of smaller proteins.}  This feature \added{is shown in
  Fig.~\ref{fig:gap_accessibility} and} may be exploited as a viable
slow-diffusion-mediated selectivity for a cell to further regulate the
accessibility of DNA to larger proteins.

\begin{figure}[h!]
  \centering
  \hspace{-1mm}
  \includegraphics[width=2.8in]{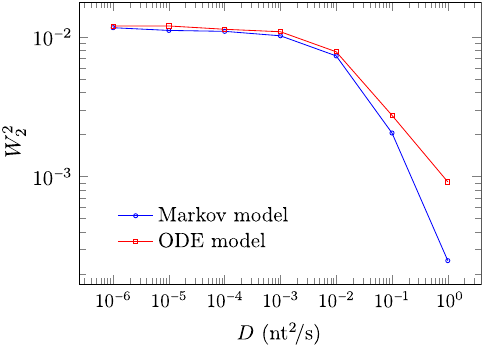}
  \caption{The squared Wasserstein distance between
    \replaced{fractional-coverage distributions}{densities} derived
    from our models and those found from full particle simulations.
    \added{The open blue circles represent the distance between
      densities predicted from the reduced Markov model described by
      Eq.~\eqref{eq:master-equation} and from agent-based
      simulations. The open red squares give} the distance between
    densities found by solving Eqs.~\eqref{eq:hydrodynamic-ode} and
    those found from agent-based simulations.  The densities derived
    from simulations are based on \added{4000} trajectories generated
    by different models \replaced{at each snapshot, followed by
      averaging the distances over the timepoints $t=0,5,\ldots,1795$
      s}{and averaging them at each timepoint}.  Here, $k_{\rm a} =
    10^{-4}$/s, $k_{\rm d} = 10^{-3}$/s, $k_{12} = 0.01$/s, $k_{21} =
    10^{-6}$/s, $a_1=20$, $a_2=30$, \added{and the system size is
      $L=5000$ nt.}}
  \label{fig:wasserstein_diffusion_compare}
\end{figure}

\added{The accuracy of the reduced models (as measured by the squared
  Wasserstein-2 distance)} can be restored by increasing the diffusion
rate $D$, as shown by Fig.~\ref{fig:wasserstein_diffusion_compare},
which compares the squared Wasserstein distance between
\replaced{distributions of the fractional-coverage vector
  $(a_1\rho_1,a_2\rho_2)$}{densities $\boldsymbol\rho(t)$} derived
from our diffusion-independent results (Eqs.~\ref{eq:master-equation},
\ref{eq:hydrodynamic-ode}) and from full stochastic
simulation\added{s} (which do depend on diffusivity $D$).  The squared
Wasserstein-2 distance decreases significantly as $D$ increases.

\subsection{\added{Application to RPA binding dynamics}}
\label{sec:rpa-binding-application}
A significant advantage of an ODE (or SDE) model over the Markov chain
model is that the former can be fitted to experimental data using
gradient-based optimization with a significantly reduced computational
cost. While some likelihood-based gradient descent methods exist for
the Markov chain model, they are not as efficient, especially when the
protein states are only partially observable.  To illustrate the
utility of our new ODE model, we revisit the experimental data in
Ref.~[\onlinecite{Ding2023June}].
\begin{figure*}
  \centering
  \includegraphics[width=5.5in]{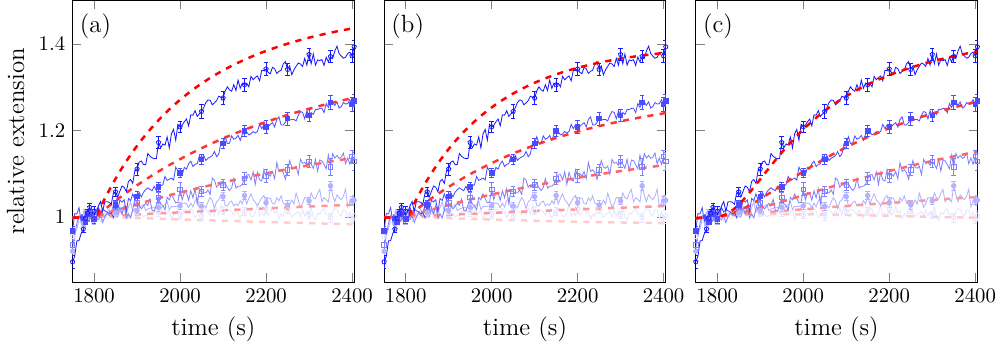}
  \caption{Fitting the ODE model to experimental data of wild-type
    RPA-DNA binding experiments under physiological salt conditions
    (150 mM NaCl). Relative ssDNA extension is measured as a function
    of time. Blue curves \added{and symbols} represent experimental
    data, \replaced{with standard-error bars, averaged over 104, 130,
      70, 86 and 45 individual strands for the 0, 1, 4, 10 and 25-fold
      conditions, respectively,}{averaged over 50 individual samples}
    while red \added{dashed} curves plot
    Eq.~\ref{eq:hydrodynamic-ode}\added{, converted to relative
      extension using Eq.~\eqref{eq:relative_extension_r34} (see
      Supplementary Information)}. Curves corresponding to each of
    five different RPA protein concentrations are distinguished by
    their darker color and higher extension levels.  (a) Using
    parameters taken from Ref.~[\onlinecite{Ding2023June}], we plot
    the corresponding model predictions (red). In (b), we performed
    fast, direct fitting of the ODE model to the measured time series
    \added{with instantaneous concentration exchange}. Here, the
    optimal parameter values clearly reduce deviations between model
    and data relative to those apparent in (a). \added{(c)
      Optimization with gradual concentration exchange,
      Eq.~\eqref{eq:concentration_exchange_r34}.}}
\label{fig:fitting_wt_150mM_salt}
\end{figure*}
\added{RPA binding to ssDNA was studied using
  DNA curtains \cite{curtains2010}, in which hundreds of DNA strands
  are aligned under flow and exposed to controlled protein
  concentrations. The relative ssDNA extension reports the dynamics of
  RPA loading. An agent-based stochastic adsorption model was fitted
  to these measurements, requiring repeated simulations to identify
  suitable parameters.}

\added{The ODE model given in Eqs.~\eqref{eq:hydrodynamic-ode} is a
  fast-diffusion mean-field approximation of the same agent-based
  binding model used in Ref.~[\onlinecite{Ding2023June}], retaining its
  attachment, detachment, and bound-mode conversion mechanisms.
  Figure~\ref{fig:fitting_wt_150mM_salt}(a) shows the ODE prediction
  with the reference parameters, and panel (b) shows direct
  optimization against the extension traces. The improved fit in panel
  (b) reflects more efficient parameter optimization, rather than a
  different physical model.}

\added{The extension traces also show a slow initial response followed
  by acceleration, particularly at the 10-fold and 25-fold
  concentrations.  The ODE formulation allows finite solution exchange
  times to be incorporated and fitted efficiently alongside the
  reaction rates. We modeled the finite time required for solution
  exchange by letting the attachment rate approach its new value
  exponentially after the concentration change, and fit the exchange
  timescale jointly with the binding parameters. The resulting fit in
  Fig.~\ref{fig:fitting_wt_150mM_salt}(c), with exchange time
  $\tau=95.46$ s, captures this delayed rise more accurately.  The
  extension observable and dynamics of concentration changes are given
  in
  Eqs.~\eqref{eq:relative_extension_r34}--\eqref{eq:concentration_exchange_r34}
  in the SI. Details of the fitting procedure and parameter
  estimates with bootstrap uncertainty ranges
  (Table~\ref{tab:r34-fit}) are also provided in the SI.}

\subsection{General multisegment protein–DNA binding dynamics}
\label{sec:facilitated-exchange}

\begin{figure*}[htb]
  \centering
  \includegraphics[width=5.5in]{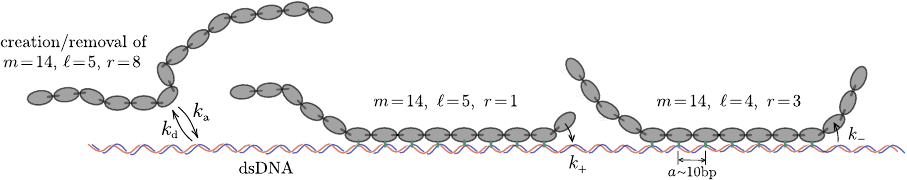}
  \caption{Schematic of multisegment protein binding. A specific
    example is histone-DNA interactions in which the contact sites
    around the histone disk are represented by a segmented protein
    with $m=14$ binding units, each occluding $a\approx 10$bp of
    DNA. The mechanism of detachment is successive peeling of segments
    from the left and right ends. These segments bind and unbind with
    rate $k_{+}$ and $k_{-}$, respectively, while protein attachment
    and detachment occur via a single segment at rates $k_{\rm a}$ and
    $k_{\rm d}$, respectively.}
  \label{fig:histone}
\end{figure*}

The kinetic models presented thus far can be straightforwardly
extended to describe binding proteins composed of a chain of multiple
\deleted{larger} segments, each covering one or more consecutive DNA
base pairs. Each bound protein can exist in a hierarchy of partially
bound configurations. An important example of this type of binding
protein is the nucleosome, whose collective occupancy on dsDNA is
\replaced{implicated in regulating access to transcription initiation
  sites and gene expression}{implicated in the accessibility of
  \textit{e.g.}, transcription initiation sites and is an important
  factor in the regulation of gene expression}
\cite{elgin1994,chen2014}. The binding, unbinding
\cite{ralf2019,Li2023November} and sliding
\cite{histones_prl,schiessel2013} of individual histones
\replaced{influence nucleosome arrangement}{mediate how nucleosomes
  are arranged}.

As depicted in Fig.~\ref{fig:histone}, a single nucleosome, when fully
wrapped by dsDNA, makes about 14 contacts that occlude about
\replaced{10~bp}{10bp} each, for a total wrapping length of $\sim$
\replaced{147~bp}{147bp}
\cite{richmond2003,Armeev2021}. \added{Histone-DNA} binding is
\added{one} natural realization \added{of binding through multiple
  protein segments}\deleted{of proteins with multiple binding
  segments,}\replaced{. More generally,}{; more generally,} consider a
\replaced{protein with symmetric ends}{directionless protein} composed
of $m$ identical binding segments, each occluding $a$ lattice sites
when bound.  A single-protein configuration is characterized by the
ordered pair $(\ell,r)$, where $\ell$ and $r$ denote the numbers of
unbound units that are peeled from the left and right ends,
respectively.  Hence, $\ell+r = m$ corresponds to a fully detached
protein, whereas $(\ell,r)=(0,0)$ represents the fully bound state.
The entire ensemble of admissible states associated with a single
protein can be represented by a triangular region of a square lattice
$\{(\ell,r)\,:\,\ell,r\ge 0,\;\ell+r\le m\}$; stochastic motion on
this lattice (nearest–neighbor jumps) captures successive unwrapping
and rewrapping of an individual protein.


\paragraph{Elementary single-protein transitions.}

Proteins first attach to DNA through binding of one of their
segments. Provided $a$ contiguous sites are open, \deleted{and
  accessible,} attachment occurs at rate $k_{\rm a}$ and generates a
single protein system in state $\ell+r=m-1$. For any state with
$\ell+r=m-1$, one unit remains bound to the DNA. Internal wrapping
transitions $(\ell,r)\to(\ell-1,r)$ or $(\ell,r)\to(\ell,r-1)$ occur
with rate $k_{+}$ when $a$ consecutive sites to the left or right of
the group of bound units are open. The reverse unwrapping steps
$(\ell,r)\to(\ell+1,r)$ or $(\ell,r)\to(\ell,r+1)$ are unconstrained
and proceed with rate $k_{-}$. These kinetic steps are depicted in
Fig.~\ref{fig:histone}. A stable, strongly bound, isolated protein
requires $k_{+} > k_{-}$. \added{In this multisegment model the last
bound segment detaches at rate $k_{\rm d}=k_{-}$. The attachment
constant $k_{\rm a}$ is the rate per accessible placement and per
entrance state; there are $m$ such states, one for each possible
initially bound segment. Retaining the fast translational
  diffusion} limit, we treat every partially peeled state as a
distinct binding mode that interconverts rapidly with its neighboring
partially peeled-state configurations before appreciable attachment or
detachment occurs.

\paragraph{Extension to many proteins.} While models of
individual histone binding and sliding
\cite{histones_prl,schiessel2013,Li2023November} and the
\textit{equilibrium} thermodynamics of ensembles of histones with steric
exclusion of contact footprints \cite{histones_2003} have been
developed, here, we construct more general and tractable models to
describe the \textit{dynamics} of bulk-exchanging multisegmented
protein binding and peeling.

Since each attached protein can take on many ($m(m+1)/2$) internal
states, a Markov master equation analogous to
Eq.~\eqref{eq:master-equation}, while straightforward to derive, is
too unwieldy to work with analytically. Thus, we will \replaced{take
the large-system-size limit directly}{directly take
the large system size limit} by defining the number density of proteins
in configuration $(\ell,r)$ by $\rho_{\ell,r}(t)$. Since each attached
protein has footprint $(m-\ell-r)a$, $0\leq \rho_{\ell, r}(t) \leq
\big[(m-\ell-r)a\big]^{-1}$ and the coverage is
\begin{equation}
  \theta \equiv \sum_{\ell, r}a (m-\ell-r)\rho_{\ell, r}.
  \label{eq:theta_rho_l_r}
\end{equation}
Since the total density is $\rho = \sum_{\ell,r}\rho_{\ell,r}$, the
accessibility parameter $\eta$ can be written as

\begin{equation}
  \begin{aligned}
    \eta \equiv &  \mfrac{1-\theta}{1-\theta + \rho}
    = 1- \frac{\sum_{\ell,r}\rho_{\ell,r}}
    {1-\sum_{\ell,r}\big[(m-\ell-r)a-1\big]\rho_{\ell,r}}.
  \end{aligned}
  \label{eq:eta_rho}
  \end{equation}
The total attachment rate into any \added{state} $(\ell,r)$ with
$\ell+r=m-1$ becomes $k_{\rm a}\eta^{a-1}(1-\theta)L$. Internal
wrapping and unwrapping events occur with \added{effective} rates
$k_{+}\eta^{a}\rho_{\ell,r}$ and $k_{-}\rho_{\ell,r}$, respectively.

\paragraph{Hydrodynamic equations.}
Collecting all contributions yields the following set of coupled
ODEs

\begin{widetext}
\begin{equation}
  \begin{aligned}
\frac{\mathrm d\rho_{\ell,r}}{\mathrm d t}
={}&\added{k_{\rm a}\eta^{a-1}(1-\theta)\delta_{\ell+r,m-1}}
+k_{+}\eta^a(\rho_{\ell+1,r}+\rho_{\ell,r+1})+k_{-}(\rho_{\ell-1,r}+\rho_{\ell,r-1})\\
&-\added{\big[k_{+}\eta^a\big(H(\ell)+H(r)\big)
+k_{-}\left(2-\delta_{\ell+r,m-1}\right)\big]\rho_{\ell,r}},
\qquad \ell,r\ge 0,\quad\ell+r < m.
\end{aligned}
  \label{ODE_multimer}
\end{equation}
\end{widetext}

\added{Here, we define $\rho_{\ell,r}=0$ for ($\ell, r$) outside the
  attached state space $\{\ell,r\geq 0, \ell + r < m\}$.  A protein
  with at least two bound segments can peel at either end, whereas
  removal of its last bound segment is a single detachment
  transition. These multiplicities also apply when $m=1$. Summing over
  attached states cancels all internal transitions and gives
\begin{equation}
\frac{\mathrm d}{\mathrm d t}\sum_{\ell+r<m}\rho_{\ell,r}
=m k_{\rm a}\eta^{a-1}(1-\theta)
-k_{-}\!\!\!\!\sum_{\ell+r=m-1}\!\!\!\rho_{\ell,r}.
\label{eq:multimer-number-balance}
\end{equation}}
Eqs.~\eqref{ODE_multimer}, through the $\rho$-dependence of $\theta$
and $\eta$ given in Eqs.~\eqref{eq:theta_rho_l_r}
and~\eqref{eq:eta_rho}, are a closed set of nonlinear ODEs describing
the dynamics of the populations of all configurations $(\ell,r)$.
They are the counterparts to Eqs.~\eqref{eq:hydrodynamic-ode} for the
binding, unbinding, and footprint changes of unoriented multisegmented
proteins.

\paragraph{\replaced{Facilitated binding through sequential segment attachment.}{Effective detachment of a single lumped segment.}}
\added{Larger footprints suppress attachment because they require
  longer contiguous gaps, as expressed by the accessibility factor
  $\eta^{a-1}$ in Eq.~\eqref{eq:hydrodynamic}. A multisegment protein
  can instead attach through a single segment of footprint $a$, rather
  than requiring the full footprint $ma$ to be vacant at once.
  Subsequent wrapping extends its contact with DNA in smaller steps as
  adjacent space becomes available. This sequential pathway allows a
  protein with a large fully bound footprint to attach more readily
  than a single-segment protein with the same total footprint.  A
  related argument was developed by Osberg et al., who showed that
  transiently reduced footprints arising from nucleosome breathing and
  stepwise assembly permit insertion into narrow gaps and accelerate
  the formation of dense nucleosome arrays \cite{Osberg2014November}.}
To illustrate how internal states shape the kinetics, we simulate a
protein with $m=3$ binding segments, each with footprint $a=2$, and
having an attachment rate $k_{\rm a}=0.5$, wrapping rate $k_{+}=1$,
and unwrapping rate $k_{-}=0.1$. For comparison, we also construct

\begin{figure}
  \centering
  \includegraphics[width=3.4in]{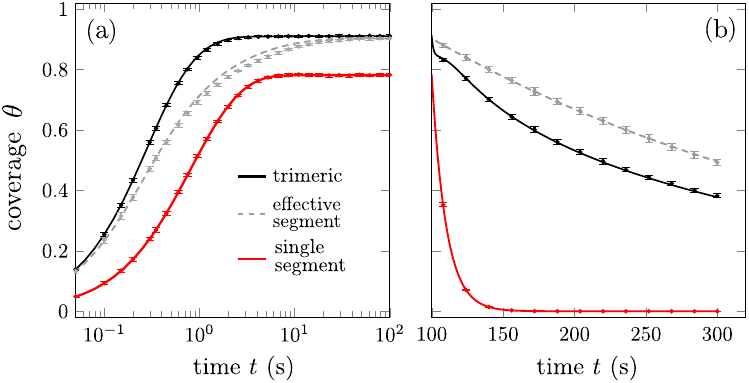}
  \caption{The dynamics of coverage by a three-segment (``trimeric'')
    protein (solid black) compared to those of two single-segment
    protein models, including an effective model in which the
    detachment rate is modified to reflect mean time to detachment of
    the full three-segment model. (a) Association phase: total
    coverage $\theta(t)$ for the three-segment model (solid
    \added{black}) using parameters $m=3$, $a=2$, $k_{\rm a}=0.5$,
    $k_{+}=1$, and $k_{-}=k_{\rm d}=0.1$, the simple single-segment
    protein model (i) (solid red) using $m=1, a=2, k_{-}=k_{\rm
      d}=0.1, k_{\rm a}=0.5$, and the effective single-segment protein
    model (ii) (\added{dashed gray}) using $m=1$, $a=6$, $k_{\rm
      a}=0.5$, \added{and $k_{\rm d}^{(1)} =0.003$}. All results were
    derived from the solutions to
    Eqs.~\eqref{ODE_multimer}. \added{The effective-segment
      approximation exhibits a slower approach to equilibrium than the
      all-segment model.}
    (b) Dissociation phase: After equilibration, $k_{\rm a}$ is set to
    zero \added{at $t=100$ s} for each model. Larger numbers of
    segments (larger $m$) result in a slower overall detachment
    process, except if modeled by an effective single-segment protein
    with decreased detachment rate $k_{\rm d}^{(1)}$. Coverages
    derived from full-lattice Monte-Carlo simulations for each model are
    shown by the symbols with error bars that indicate $\pm$ one
    standard deviation over 24 sample trajectories \added{starting
      from an empty lattice}. For these simulations, the system size
    was taken to be $L=10,000$ \added{with periodic boundaries and a
      large diffusivity $D=10$ nt$^{2}$/s was imposed.}}
  \label{fig:multi_histone}
\end{figure}

\begin{itemize}

  \item[(i)] a simple single-segment ($m=1$) protein with
  $a=2$, $k_{\rm a}=0.5$, $k_{\rm d}=0.1$, and

  \item[(ii)] a ``lumped'' protein composed of a single
    \emph{effective segment} with footprint $a_\mathrm{eff}=6$
    that retains the same $k_{\rm a}$ but has \replaced{an effective
    detachment rate}{an effective peeling and detachment rates}
    $k_{\rm d}^{(1)}=mk_{+}(k_{-}/k_{+})^{m}$
    \cite{Li2023November} chosen so that its mean detachment time
    \replaced{approximates}{matches} that of the full three-segment protein.
\end{itemize}
Figure \ref{fig:multi_histone} plots the coverage associated with the
numerical solutions to Eqs.~\eqref{ODE_multimer} using parameters
corresponding to the dynamics of the full three-segment protein and
comparison cases (i) and (ii). The trimeric protein and the
single-effective-segment protein with effective detachment rate
$k_{\rm d}^{(1)}$ possess comparable times to detachment and reach
nearly identical equilibrium coverages $\theta$; however, the true
multisegment protein approaches the equilibrium coverage appreciably
faster because its initial attachment requires open gap sizes of only
$a=2$.

\paragraph{Facilitated exchange between two protein species.}
We now introduce two protein types that attach to the same lattice,
have the same physical features and rates, and compete for the same
global vacancy $\eta=(1-\theta)/(1-\theta+\rho)$ ($\rho$ is the total
density of bound proteins). \added{Such competition can accelerate the
  replacement of one protein by another and is an important mechanism
  in, \textit{e.g.}, transcription factor binding \cite{marko2017}.}
\added{A theoretical analysis comparing competitive binding of two
  different-sized particles in the fast diffusion (equilibrated) and
  no-diffusion limits is given in Ref.~[\onlinecite{GAP}]. Here, the
  evolution of coverage depends subtlely on the gap distributions when
  the second protein is introduced, suggesting how interparticle
  binding competition can influence global distributions and dynamics.
  Cocco et al. described a stochastic-ratchet mechanism of facilitated
  exchange: partial release of an attached protein exposes sites to a
  protein from solution, whose binding blocks reattachment of segments
  of the partially bound incumbent protein \cite{cocco2014}.  Sing et
  al. demonstrated this mechanism in Brownian-dynamics simulations of
  many dimers on a DNA chain and developed a reduced kinetic model for
  their exchange \cite{sing2014}.  Our earlier single-nucleosome model
  characterized facilitated histone detachment when competing proteins
  occupy DNA contacts exposed by partial peeling
  \cite{Li2023November}.}

\added{The present framework brings these facilitated-detachment and
  exchange mechanisms into a common description of collective protein
  binding dynamics in which proteins with an arbitrary number of
  segments are described by coupled population equations for their
  partially bound states.  Here, we define distinguishable ``type 1''
  and ``type 2'' proteins that are different only in identity; they
  carry identical sizes $m$ and kinetic rates. During time $0 \leq t
  \leq 20$, only type-1 proteins are present (and available to attach
  with rate $k_{\rm a} = 0.5$). For $t=20$, we replace the bulk
  solution with type-2 proteins that carry identical microscopic
  parameters.}


\added{Under fast lateral diffusion, the shared accessibility $\eta$
  couples attachment and rewrapping to the evolving protein
  populations, so facilitated exchange emerges without an additional
  exchange rule.  Moreover, when the total coverage $\theta$ is slowly
  changing, $\eta$ remains approximately constant, effectively
  renormalizing the internal wrapping rate to $k_{+}\eta^{a} \equiv
  \widetilde{k}_{+}$.}  An $m$-segment protein can also be
approximated by a single-segment protein with an effective,
environment-dependent detachment rate

\begin{equation}
  \begin{aligned}
    k_{\rm d}^{\rm eff}(\eta) & \approx k_{\rm d}^{(1)}\added{\eta^{-a(m-1)}} =
    mk_{+}\eta^{a} \bigg(\frac{k_{-}}{k_{+}}\bigg)^{m}\eta^{-am}\\
  \: & = m\,\added{\widetilde k_{+}}
  \bigg(\frac{k_{-}}{\added{\widetilde k_{+}}}\bigg)^m,
  \end{aligned}
  \label{eq:facilitated_unbinding}
\end{equation}
analogous to the single-protein result discussed in
Ref.~[\onlinecite{Li2023November}]. \added{This is a strong-rewrapping asymptotic
  estimate, requiring $\widetilde{k}_{+}\gg k_{-}$. We use the
  switch function}

  \begin{equation}
  \added{k_{\rm d}^{\rm eff} \sim
    \min\Big\{m\widetilde{k}_{+}\Big(\mfrac{k_{-}}
      {\widetilde{k}_{+}}\Big)^m,\,\mfrac{2k_{-}}{m+1}\Big\}}
  \label{eq:interpolation}
  \end{equation}
\added{to impose a limit to the detachment rate when $\eta$ becomes
  too small and $\widetilde{k}_{+} \not\gg k_{-}$. The pure-peeling,
  no rewrapping detachment rate is heuristically set to $2k_{-}/(m+1)$.}
The relative exchange rate can then be approximated by
$\mathrm{d}\theta_{1}(t)/\mathrm{d} t = -k_{\rm d}^{\rm
  eff}(\eta)\theta_{1}(t)$, where $\eta$ can be evaluated at its
steady-state fully loaded value.  This approximation yields
$\theta_{1}(t)=\theta_{1}(0)\exp[-k_{\rm d}^{\rm eff}(\eta)t]$.
  
\begin{figure}
  \centering
  \includegraphics[width=3.4in]{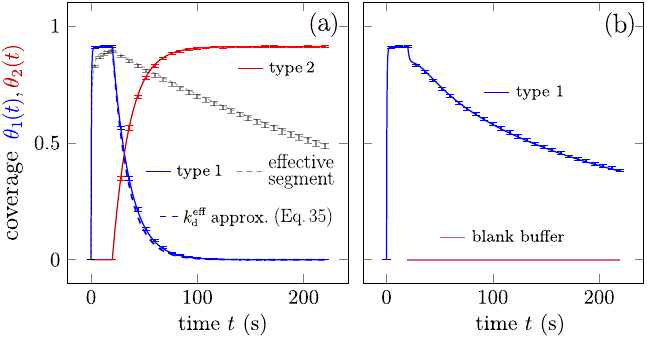}
  \caption{Facilitated exchange experiment. (a) \added{Coverage of
      type-1 proteins found} by solving Eq.~\ref{ODE_multimer} using
    parameter values $m=3, a=2, k_{+}=1, k_{\rm a} = 0.5$, and
    $k_{-}=0.1$. \added{Substituting the solution $\rho_{\ell, r}$
      into Eq.~\ref{eq:theta_rho_l_r}, we find and plot the type-1
      protein coverage $\theta_{1}(t)$.} After type-1 proteins (solid
    blue curve) have nearly equilibrated on the whole lattice, the
    solution is replaced with type-2 proteins at $t=20$ (solid red
    curve). This is achieved by setting $k_{\rm a} =0$ for type-1
    molecules while turning on $k_{\rm a}=0.5$ for type-2
    molecules. \replaced{The time-dependent coverages}{The
      time-dependence of the coverages} $\theta_{1}(t)$ and
    $\theta_{2}(t)$ show that replacement of type-1 by type-2 protein
    is much faster than what an effective, lumped single-segment
    approximation model predicts (dashed gray curve). The dashed blue
    curve is the analytic approximation $\theta_{1}(t\geq
    20)=\theta_{1}(20)\exp[-k_{\rm d}^{\rm eff}(\eta)(t-20)]$ using
    $k_{\rm d}^{\rm eff}(\eta)$ defined in
    \added{Eq.~\eqref{eq:interpolation}.} (b) If the type-1 protein
    solution is removed at $t=20$ without replacement (blank
    solution), its \replaced{clearance is slower without incoming
      competitors}{removal is no longer incrementally facilitated,
      leading to a much slower decay of $\theta_{1}(t)$}.}
  \label{fig:facilitated_exchange}
\end{figure}

Fig.~\ref{fig:facilitated_exchange}(a) shows that during substrate
loading ($t\leq 20$), the effective-segment model (dashed gray,
comparison model (ii) above) provides a good approximation to the
coverage $\theta$ obtained by numerically solving
Eq.~\eqref{ODE_multimer} (solid blue curve). After $t>20$, the full
multisegment ($m=3$) model reveals rapid displacement of pre-bound
type-1 proteins. \replaced{This facilitated exchange arises from
  incremental peeling and competitor invasion, a pathway absent from
  the fixed-rate single-segment comparison model, which predicts much
  slower protein clearance.}{This \emph{facilitated exchange} behavior
  arising from incremental peeling is not captured by any
  single-segment model, which predicts much slower protein clearance.}
For $m>1$, incoming proteins occupy vacancies created by partial
peeling, preventing rewrapping and thereby biasing the incumbent
protein toward full detachment. \added{Resolving this pathway provides
  a mechanistic basis for a competitor-dependent effective detachment
  rate, as in the multiple-binding-site model of Sing et al.
  \cite{sing2014}.} Fig.~\ref{fig:facilitated_exchange}(b) shows the
removal of the type-1 protein solution at $t=20$ \textit{without}
replacing it with type-2 protein. In the absence of a competitive
binder that can facilitate peeling, protein coverage decreases much
more slowly.  These results demonstrate that our multistep peeling
model is able to capture the essential \added{competitive} feature of
facilitated exchange, despite having no explicit interactions
particular to type-1 and type-2 proteins other than hard steric
exclusion.  Experimentally, similar facilitated exchange of RPA
proteins has been observed using DNA curtains \cite{Gibb2014February}.

\section{Discussion \& Conclusions}
\label{sec:discussion}
The dynamic interplay between DNA-binding proteins and their
substrates is central to genomic regulation and repair. By developing
a reduced stochastic framework for protein-DNA interactions in the
fast diffusion limit, our results, summarized by
Eqs.~\eqref{eq:master-equation}, \eqref{eq:hydrodynamic-ode}, and
\eqref{ODE_multimer}, bridge a critical gap between microscopic and
tractable macroscopic models.

By coarse-graining positional degrees of freedom under the fast
diffusion assumption, we derived an effective Markov chain model and
its associated mean-field hydrodynamic ODE.  \replaced{These reduced
  models approximate the full-lattice kinetics in the tested
  fast-diffusion regimes and enable direct fitting of experimental
  extension curves without resolving individual protein
  positions.}{These reduced models quantitatively replicate the
  dynamics of full agent-based simulations and experimental data while
  drastically lowering computational costs.} The validity of the fast
diffusion approximation hinges on the separation of timescales between
rapid positional equilibration and slower binding/unbinding or
conformational transitions, assessed by the local density-dependent
comparison and its conservative envelope described by
Eq.~\eqref{eq:practical-diffusion-region}.  This framework
successfully predicts how protein footprint size reduces site
accessibility exponentially \added{through the factor $\eta^{a-1}$,
  even at fixed total coverage,} as found previously in 1D Tonks gas
models.

Additionally, the model reveals intrinsic limitations of DNA
accessibility: larger footprints limit available sites but
multisegmented structures mitigate this penalty. By adopting
hierarchical, partially bound states, multisegmented proteins like RPA
can use smaller segments to accelerate initial attachment while
retaining high overall affinity. \added{This facilitated binding
  requires only a small gap for initial attachment, followed by
  gradual engagement of additional segments as space becomes
  available.  Related variable-footprint models show that nucleosome
  breathing and stepwise assembly accelerate the formation of dense
  arrays \cite{Osberg2014November}. Partial binding also implements
  the stochastic-ratchet mechanism of facilitated exchange
  \cite{cocco2014,sing2014}, in which incoming proteins block
  rebinding of partially released incumbents
  \cite{chen2014,luo2014}. The same competition promotes complete
  detachment, connecting exchange to facilitated unbinding
  characterized in our earlier single-nucleosome model
  \cite{Li2023November}. The present population equations bring these
  binding, exchange, and unbinding pathways into a common description
  of multisegment proteins.} Such dynamics align with experimental
observations of competitive protein turnover on ssDNA, which
illustrate the biological relevance of conformational plasticity.

While our model provides insights into 1D diffusion-driven binding, it
can be extended to other contexts.  For example, ATP-dependent
remodeling factors, such as helicases or chromatin remodelers,
introduce energy-driven processes that alter binding
equilibria. Incorporating nonequilibrium molecular motors into the
model could capture active displacement mechanisms within
nonequilibrium steady states, applicable to replication and repair
mechanisms. We expect our approach \added{to be applicable} to driven
diffusive systems such as asymmetric exclusion processes involving
Langmuir adsorption kinetics \cite{frey_langmuir_2004} and
variable-size particles \cite{kern2019,dropoff2022}.
  
Another important class of DNA-binding proteins consists of proteins
that bind specifically to target DNA sequences, while maintaining a
general nonspecific affinity for the rest of the DNA. A hybrid
framework combining our fast-diffusion formalism with slow or immobile
components could elucidate how stable ``bookmarking'' proteins
influence the dynamics of mobile regulators. A heterogeneous substrate
would require multiscale methods that retain finite diffusivity near
target sites that could be reflected in boundary conditions of
hydrodynamic \textit{PDE} models for adsorbed protein density.
Finally, our model for multisegmented protein adsorption assumed fast
translational diffusion relative to internal segment binding/unbinding
kinetics, which \added{may often} not hold. For example, histone
sliding is thought to be mediated by multiple segment unbinding and
rebinding steps \cite{histones_prl} or twist defects
\cite{schiessel2003} that travel through the contact footprint. The
regime in which internal state kinetics are fast compared to
translational diffusion is another natural limit open for further
analysis.

\added{The preceding reduced-model analysis assumes fast diffusion.
Our detailed-balance argument shows that diffusion does not alter the
equilibrium distribution, but insufficient diffusion can lead to slow
spatial relaxation when the overall unbinding process is slow.}

\added{Slow spatial relaxation can keep DNA--protein assemblies out of
  equilibrium over biologically relevant
  timescales. Fig.~\ref{fig:supp-contraction} in the SI shows
  that a clumped initial configuration can relax slowly without
  diffusion when conversion from mode 2 to mode 1 limits subsequent
  detachment and rebinding. If this relaxation is slower than protein
  turnover or changes in the cellular environment, the system may
  remain out of equilibrium throughout the relevant biological
  process, making transient kinetics more informative than equilibrium
  occupancy alone.}



\clearpage

\setcounter{figure}{0}
\setcounter{equation}{0}
\setcounter{table}{0}
\setcounter{page}{1}
\renewcommand{\theequation}{S\arabic{equation}}
\renewcommand{\thetable}{S\arabic{table}}
\renewcommand{\thefigure}{S\arabic{figure}}
\onecolumngrid
\section*{\added{Supplementary Information}}

\begin{figure}[H]
\centering
\begin{minipage}{\textwidth}
  \centering
  \includegraphics[width=0.88\textwidth]{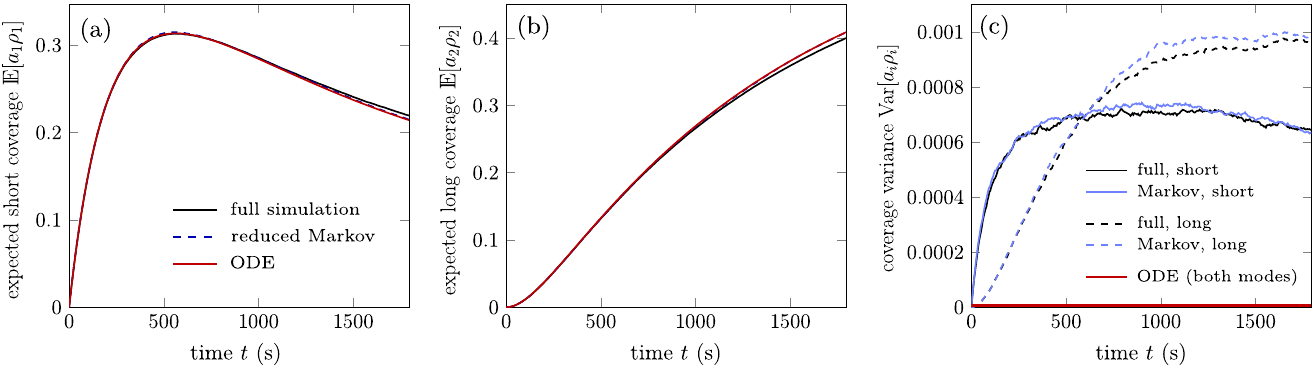}
\caption{\added{Mean and variance of the coverage trajectories.
    Panels (a) and (b) show the ensemble means of the short-mode
    coverage $a_1\rho_1$ and long-mode coverage $a_2\rho_2$, where
    $\rho_i=N_i/L$ is the number density defined in the main text.
    Panel (c) shows $\mathrm{Var}(a_i\rho_i)$ across trajectories for
    $i=1,2$. Black curves denote the full simulation and blue curves
    denote the reduced Markov model. In (a) and (b), the reduced
    Markov means are dashed and the ODE predictions are red. In (c),
    solid and dashed curves denote short- and long-mode variances,
    respectively.}}
\label{fig:supp-mean-variance}
\end{minipage}
\end{figure}

\noindent {\bf \textit{\added{Fluctuations in reduced-model dynamics.}}}
\added{Figure~\ref{fig:supp-mean-variance} compares ensemble means and sample
  variances between the stochastic models, using 4000 independent
  trajectories per stochastic model.  Both ensembles start from an
  empty lattice with $L=5000$ nt, $a_1=20$ nt, $a_2=30$ nt, $D=0.1$
  nt$^2$/s, $k_{\rm a}=10^{-4}$, $k_{\rm d}=10^{-3}$,
  $k_{12}=10^{-3}$, and $k_{21}=10^{-4}$ s$^{-1}$.  The mode coverages
  $a_iN_i/L$ are evaluated every 5 s over 0--1795 s. The three models
  have closely matching mean coverages, while the reduced Markov model
  also reproduces the nonzero variances of the full spatial
  model. This accounts for its smaller squared Wasserstein distance at
  these parameters.}
%

\noindent {\bf \textit{\added{Reaction-mediated redistribution.}}}
\added{Here, we compare systems with and without diffusion to determine how
attachment and detachment alone redistribute proteins. The full
spatial model uses $a_1=20$ nt and $a_2=30$ nt on a closed lattice of
$L=200$ nt.  Each ensemble contains 1000 independent simulated
trajectories initialized with two proteins in each mode packed
together at one end, with their mode order randomized. We compare the
coverages $\theta=(a_1N_1+a_2N_2)/L$ in the $D=0$ and $D=10$ nt$^2$/s
cases, with all other rates identical.  The thin dashed curves show
the ensemble mean plus or minus 1.96 standard errors.  The equilibrium
coverage is calculated from the distribution defined in the subsection
``Equilibrium with and without diffusion'' by summing over admissible
counts. For given $N_1,N_2$, the number of configurations is
$\binom{N_1+N_2}{N_1}\binom{R+N_1+N_2}{N_1+N_2}$, where
$R=L-a_1N_1-a_2N_2$.}
%
\begin{figure}[H]
  \centering
  \begin{minipage}{\textwidth}
    \centering
    \includegraphics[width=0.88\textwidth]{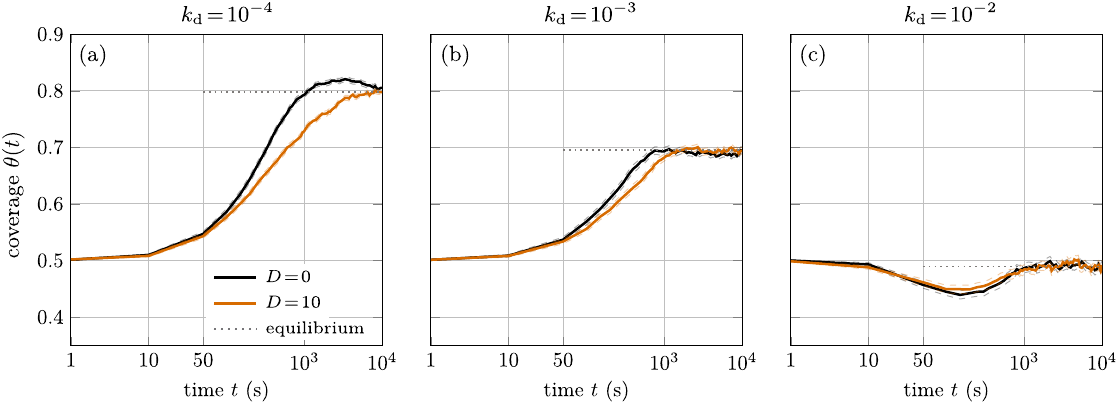}
\caption{\added{Reaction-mediated relaxation for increasing detachment
    rates. Panels (a), (b), and (c) use $k_{\rm
      d}=10^{-4},10^{-3},10^{-2}$ s$^{-1}$, respectively, with
    $L=200$, $k_{\rm a}=10^{-4}$, $k_{12}=10^{-2}$, and
    $k_{21}=10^{-3}$ s$^{-1}$. Each ensemble contains 1000
    trajectories starting from two proteins in each mode clumped
    together. Black and orange curves show mean coverage at $D=0$ and
    $D=10$, respectively. Thin dashed curves show mean $\pm1.96$
    standard errors, and horizontal dotted lines show exact
    equilibrium coverage.}}
\label{fig:supp-unbinding}
\end{minipage}
  \end{figure}

\begin{figure}[H]
  \centering
\begin{minipage}{\textwidth}
    \centering
\includegraphics[width=0.88\textwidth]{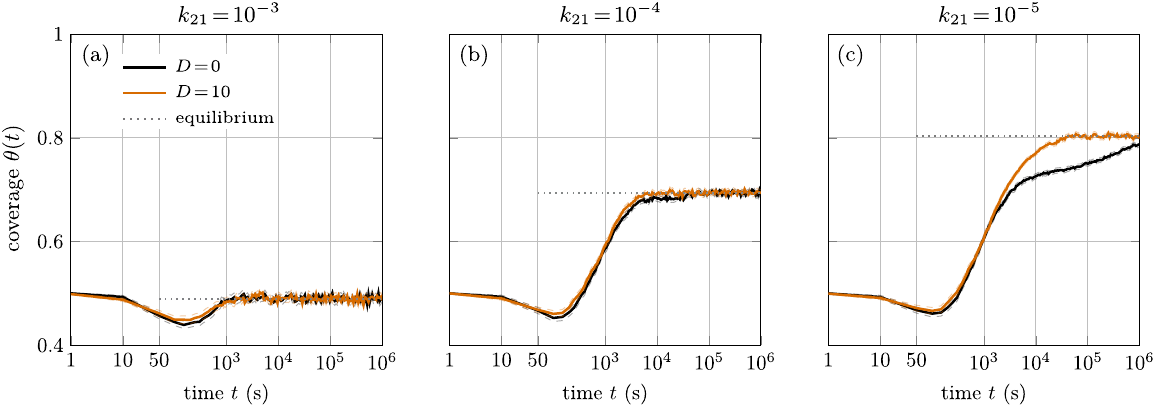}
\caption{\added{Reaction-mediated relaxation at different contraction
    rates. Each panel compares $D=0$ and $D=10$ at $L=200$, $k_{\rm
      a}=10^{-4}$, $k_{\rm d}=10^{-2}$ and $k_{12}=10^{-2}$ s$^{-1}$,
    with $k_{21}=10^{-3},10^{-4},10^{-5}$ s$^{-1}$ in panels (a), (b),
    (c), respectively. Each ensemble contains 1000 trajectories,
    initially with two proteins in each mode clumped together. Black
    and orange curves represent mean coverages while the thin dashed
    curves show mean $\pm 1.96$ standard errors.  Horizontal reference
    levels (dotted line segments) show exact equilibrium coverage
    values for each fixed-rate pair.}}
\label{fig:supp-contraction}
\end{minipage}
\end{figure}

\added{In Fig.~\ref{fig:supp-unbinding}, we fix $L=200$, $k_{\rm
    a}=10^{-4}$, $k_{12}=10^{-2}$, and $k_{21}=10^{-3}$ s$^{-1}$ and
  scan $k_{\rm d}=10^{-4},10^{-3}$, and $10^{-2}$ s$^{-1}$ across (a),
  (b), and (c), respectively. In Fig.~\ref{fig:supp-contraction}, the
  detachment rate $k_{\rm d}=10^{-2}$ s$^{-1}$ is fixed, but the
  conversion rate $k_{21}$ is varied from $10^{-3}$ to $10^{-5}$
  s$^{-1}$ across the panels. The observation windows extend to $10^4$
  and $10^6$ s, respectively. Trajectories were recorded every 50 s.
  The points at 1 and 10 s were obtained from separate ensembles with
  the same initial-condition distribution and parameters.}

\added{Faster detachment reduces the difference between the diffusive
  and diffusionless trajectories in Fig.~\ref{fig:supp-unbinding}.
  When detachment is slow, the diffusionless system transiently exceeds its
  equilibrium coverage before relaxing. A protein in mode 2 must
  convert to mode 1 before detachment can redistribute
  it. Consequently, slow conversion creates a prolonged intermediate
  plateau without diffusion in Fig.~\ref{fig:supp-contraction}(c). The
  common equilibrium therefore does not imply comparable relaxation
  times from a spatially clumped configuration.}

\begin{figure}
\centering
\includegraphics[width=4.4in]{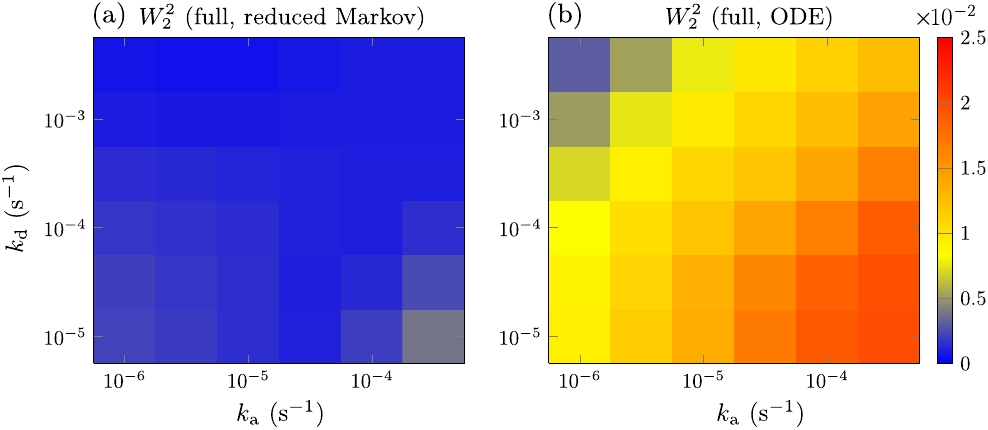}
\caption{\added{Error across distributions predicted by different
    models plotted across attachment and detachment rates. (a)
    compares the full model to the reduced Markov model while (b)
    compares it to the ODE model. The fixed parameters are $L=500$ nt,
    $D=0.1$ nt$^{2}$/s, $k_{12}=10^{-3}$ s$^{-1}$, and
    $k_{21}=10^{-4}$ s$^{-1}$, with 1000 trajectories per stochastic
    ensemble and two proteins initially in each mode.  Squared
    Wasserstein distances use fractional-coverage coordinates and are
    averaged at 5 s intervals over 0--10000 s.}}
\label{fig:supp-binding}
\end{figure}
\added{Fig.~\ref{fig:supp-binding} tests the accuracy of reduced
  models across different attachment--detachment rate values. We use
  $L=500$ nt, $D=0.1$ nt$^2$/s, $k_{12}=10^{-3}$ and $k_{21}=10^{-4}$
  s$^{-1}$, with two proteins initially in each mode.  Each rate spans
  six logarithmically spaced values, with $10^{-6}\le k_{\rm
    a}\le10^{-3.5}$ and $10^{-5}\le k_{\rm d}\le10^{-2.5}$ s$^{-1}$.
  For each combination, 1000 trajectories were simulated for both the
  full and reduced Markov models, using rates under the hydrodynamic
  approximations in Eq.~\eqref{eq:hydrodynamic}. The corresponding ODE
  starts from the same protein densities. Squared Wasserstein
  distances are calculated in $(a_1N_1/L,a_2N_2/L)$ coordinates every
  5 s over 0--10000 s and averaged over time.}

\added{Faster detachment generally improves agreement by allowing
  repeated attachment and detachment to rearrange the gaps. Slow
  detachment with rapid attachment instead preserves
  adsorption-generated spatial biases.}

\newpage
\noindent {\bf \textit{\added{Parameter estimation from DNA-extension
      measurements.}}} \added{We reanalyze the 150 mM NaCl DNA-curtain
  data of Ref.~\cite{Ding2023June}. The measured extension is obtained
  by weighting free DNA and the two bound modes by their extensions
  per nucleotide. With $N_i$ proteins in mode $i$ on a strand of $L$
  nucleotides, $\rho_i=N_i/L$ is the protein number density and
  $\theta_i=a_i\rho_i$ is the nucleotide coverage, where $a_1=20$ nt
  and $a_2=30$ nt. If, relative to free DNA, the extensions per
  nucleotide are defined as $\beta$ for the 20-nt mode and $\alpha$
  for the 30-nt mode, the coverage-weighted extension and its
  pre-exchange normalization are}
\begin{equation}
  \added{E(t) =\beta\theta_1(t)+\alpha\theta_2(t)+1-\theta_1(t)-\theta_2(t)
  =1+(\beta-1)a_1\rho_1(t)+(\alpha-1)a_2\rho_2(t), \,\,\,\,
  R(t)=\frac{E(t)}{\overline E_{\rm pre}}.}
 \label{eq:relative_extension_r34}
\end{equation}
\added{Here, $E(t)$ is the extension relative to naked DNA of the same
  length, and $\overline E_{\rm pre}$ is the mean over four
  pre-exchange reference samples. Each experimental strand is
  normalized by its own corresponding reference mean before computing
  condition means and standard errors. We fix $\alpha=1.5$ and
  $\beta=2.4$ at their 150 mM values and fit the reaction rates within
  reference-centered ranges, retaining $k_{12}>k_{21}$. The rate
  $k_{12}$ describes expansion from 20 to 30 nt and $k_{21}$ describes
  contraction.}

\added{Both fits use the five concentration folds
$f\in\{0,1,4,10,25\}$ and minimize the summed mean-squared residual
over 1750--2405 s. Squared residuals from 2220 s onward receive weight
3 to emphasize the later dynamics. Direct optimization uses an
instantaneous concentration change. To represent finite solution
exchange, we instead let the attachment rate approach its new value
exponentially,}
\begin{equation}
 \added{k_{\rm a} \Rightarrow  k_{\rm a}\left[1+(f-1)
 \left(1-e^{-(t-t_0)/\tau}\right)\right],\quad t\ge t_0,}
 \label{eq:concentration_exchange_r34}
\end{equation}
\added{where $t_0=1800$ s and $\tau$ is shared across conditions.
  Before exchange, the attachment rate is $k_{\rm a}$. For $f>1$, the
  attachment rate initially remains near its pretreatment value and
  increases as the new solution arrives. The weighted squared residual
  decreases from 0.00236 for the direct fit to 0.000805 with gradual
  exchange, a reduction of 66\%.}
%
\begin{table*}[h!]
\centering
\begin{minipage}{\textwidth}
  \centering
  \caption{\added{Reference and fitted parameters for the 150 mM NaCl
      extension comparison in Fig.~\ref{fig:fitting_wt_150mM_salt}.
      Rates are in s$^{-1}$ and time is measured in seconds. The
      bootstrap ranges use a wider search than the point estimates,
      with the extension factors fixed as stated in the text.}}
\label{tab:r34-fit}
\begin{ruledtabular}
\begin{tabular}{lcccc}
\added{Parameter} & \added{Reference (a)} & \added{Direct fit (b)} &
\added{Gradual exchange (c)} & \added{95\% bootstrap range}\\
\hline
\added{$k_{\rm a}$} & \added{$5.62\times10^{-6}$} & \added{$5.51\times10^{-6}$} & \added{$1.15\times10^{-5}$} & \added{$[7.74\times10^{-6},1.40\times10^{-5}]$}\\
\added{$k_{\rm d}$} & \added{$4.22\times10^{-5}$} & \added{$1.41\times10^{-5}$} & \added{$1.41\times10^{-5}$} & \added{$[0.843,5.67]\times10^{-5}$}\\
\added{$k_{12}$} & \added{$7.50\times10^{-4}$} & \added{$1.98\times10^{-3}$} & \added{$2.50\times10^{-4}$} & \added{$[1.50,6.71]\times10^{-4}$}\\
\added{$k_{21}$} & \added{$1.78\times10^{-5}$} & \added{$5.33\times10^{-5}$} & \added{$5.33\times10^{-5}$} & \added{$[0.356,8.89]\times10^{-5}$}\\
\added{$\tau$ (s)} & \added{0} & \added{0} & \added{95.46} & \added{$[64.5,122.1]$}
\end{tabular}
\end{ruledtabular}
\end{minipage}
\end{table*}

\added{Table~\ref{tab:r34-fit} gives the reference and fitted
  parameters and the uncertainty ranges for the gradual-exchange
  fit. The ranges are the 2.5th and 97.5th percentiles from 200
  whole-strand bootstrap resamples, drawn independently within each
  condition. Each resample is refitted with the same rate ordering and
  fixed extension factors, but wider reference-centered rate ranges.
  The ranges are conditional on these assumptions and the imposed
  bounds. Expansion and detachment reach their lower bounds in 20\%
  and 93\% of resamples, respectively, and contraction reaches its
  upper bound in 88\%. The individual reaction-rate ranges therefore
  remain substantially affected by the constraints.}
%

\end{document}